\documentclass[twocolumn]{aastex63}
\usepackage{stfloats}
\usepackage{float}
\usepackage{graphicx}
\usepackage{natbib}
\usepackage{hyperref}
\usepackage{amsmath}
\usepackage{autobreak}
\usepackage{multirow}
\usepackage{makecell}
\usepackage{color}
\usepackage{bm}
\usepackage{appendix}
\usepackage{threeparttable}
\usepackage[figuresright]{rotating}

\received{}
\revised{}
\accepted{}
\submitjournal{ApJ}

\shorttitle{Binary fractions of GK dwarfs: impacts of the chemical abundances} 
\shortauthors{Niu et al.}

\graphicspath{{./}{}}

\begin{document}

\title{Binary fractions of G and K dwarf stars based on the Gaia EDR3 and LAMOST DR5: impacts of the chemical abundances} 

\correspondingauthor{Haibo Yuan}
\email{yuanhb@bnu.edu.cn}

\author{Zexi Niu}
\affiliation{National Astronomical Observatories,
Chinese Academy of Sciences\\
20A Datun Road, Chaoyang District,
Beijing, China}

\author{Haibo Yuan}
\affiliation{Department of Astronomy,
Beijing Normal University \\
19th Xinjiekouwai Street, Haidian District,
Beijing, China}

\author{Song Wang}
\affiliation{National Astronomical Observatories,
Chinese Academy of Sciences\\
20A Datun Road, Chaoyang District,
Beijing, China}

\author{Jifeng Liu}
\affiliation{National Astronomical Observatories,
Chinese Academy of Sciences\\
20A Datun Road, Chaoyang District,
Beijing, China}

\begin{abstract}

Basing on the large volume \textit{Gaia} Early Data Release 3 and LAMOST Data Release 5 data, we estimate the bias-corrected binary fractions of the field late G and early K dwarfs.
A stellar locus outlier method is used in this work, which works well for binaries of various periods 
and inclination angles with single epoch data. 
With a well-selected, distance-limited sample of about 90 thousand GK dwarfs covering wide stellar chemical abundances, 
it enables us to explore the binary fraction variations with different stellar populations. 
The average binary fraction is 0.42$\pm$0.01 for the whole sample.
Thin disk stars are found to have a binary fraction of 0.39$\pm$0.02, thick disk stars own a higher one of 0.49$\pm$0.02, while inner halo stars possibly own the highest binary fraction. 
For both the thin and thick disk stars, the binary fractions decrease toward higher [Fe/H], [$\alpha$/H], and [M/H] abundances. However, the suppressing impacts of the [Fe/H], [$\alpha$/H], and [M/H] are more significant for the thin disk stars than those for the thick disk stars. 
For a given [Fe/H], a positive correlation between [$\alpha$/Fe] and the binary fraction is found for the thin disk stars. However, this tendency disappears for the thick disk stars. 
We suspect that it is likely related to the different formation histories of the thin and thick disks. 
Our results provide new clues for theoretical works on binary formation.

\end{abstract}


\section{Introduction} \label{sec:intro}

Binary systems are ubiquitous in our universe.
Understanding star formation and evolution requires a comprehensive knowledge of binary systems. Being progenitors and hosts of various interesting systems, they are important in many fields, such as Type Ia supernovae (e.g., \citealp{2000iasn}), short-duration Gamma-Ray Bursts (e.g., \citealp{1992grb}), kilonova (e.g., \citealp{2017kilonova}), cataclysmic variables (e.g., \citealp{1995cv}), X-ray binaries, stellar population synthesis (e.g., \citealp{2002sps}), and gravitational wave astronomy (e.g.,  \citealp{2017gw170817}).

In the past few decades, the statistics of binary systems have been extensively studied with various observational methods: visual techniques \citep{1989visual,1996visual}, radial velocity variation \citep{gao2014,gao2017,tian2018,2018badnes}, spectral fitting \citep{2007wdms,2018specbi,2020pyhammer,2021xiang}, eclipsing binary \citep{2011keplereb1,2011keplereb2,2019tesseb}, common proper motion \citep{2004cpm,2020Hartmancpm,widedr2}, displacement in the color-magnitude diagram \citep{liuchao2019,2012Milone}, astrometric noise excess \citep{2019Kervella,2020Penoyre}, and color locus outlier \citep{2015loci2}. Taking advantage of multiple methods and results above, some review works summarize and present censuses of binary systems in the solar neighborhood  such as \citet{raghavan2010}, \citet{moe2017}, and \citet{moe2019anti}.
The positive correlation between the binary fraction and stellar mass is well established as well as the anti-correlation between the close binary fraction and metallicity. 

However, works focusing on the variations of the binary fraction with $\alpha$ elements are rare. The reason includes not only the relatively small data volume but also the bias of the individual method in period, mass ratio, and distance limitations, etc \citep{moe2017}. To overcome the biases and limitations of previous techniques, \citet{2015loci2} proposed a stellar locus outlier (SLOT) method to estimate model-free binary fraction for large numbers of stars of different populations in large survey volumes. 
Applying the method to about 10,000 stars from the Sloan Digital Sky Survey (SDSS) Stripe 82, \citet{2015loci2} have determined the binary fractions of different stellar populations and analyzed their dependency on spectral type and [Fe/H]. 
They find the highest binary fraction in the Galactic halo and comparable values in the thin and thick disks. 
The result is consistent with \citet{moe2019anti}, but against those from \citet{1983Carney} and \citet{2002Latham}. 
For $\alpha$ abundances, \citet{tian2018} suggest a positive correlation between [$\alpha$/Fe] and binary fraction for solar-type stars in the thin disk. 
Recently, \citet{aopgee2020alpha} perform a detailed study on the close binary fraction as a function of stellar parameters including $\alpha$ abundances and conclude that $\alpha$ elements suppress multiplicity at most values of [Fe/H]. 
More efforts are needed to explore the variations of the binary fraction with $\alpha$ elements.

Thanks to the unprecedented data volume and quality of \textit{Gaia} \citep{gaia2016} and LAMOST \citep{zhao2012} surveys, we can apply the SLOT method to a more advanced sample in terms of larger sample size and better data quality. 
The LAMOST-\textit{Gaia} stars serve as an adequate sample to explore the binary fractions with various factors, especially [$\alpha$/Fe] and stellar populations.

The paper is organized as follows. Section \ref{sec:data} describes the data selection, mainly focusing on late G and early K dwarfs in the solar vicinity. The SLOT method is described in Section \ref{sec:method}. 
Section \ref{sec:result} presents the results for the individual samples and explores the impacts of stellar parameters on binary factions. 
The results are discussed in Section \ref{sec:dis} and compared with previous works. Section \ref{sec:sum} concludes by summarizing the main findings and discussing future work.

\section{Data} \label{sec:data}

In this work, we use the \textit{Gaia} Early Data Release 3 (EDR3; \citealp{gaia2020}) and the fifth data release of LAMOST (LAMOST DR5; \citealp{zhao2012,luo2012}), accompanying with the $\alpha$ abundances provided by \citet{xiangdd}. \textit{Gaia} EDR3 provides 1.8 billion photometric data of $G$ band and 1.5 billion of $G_{\rm BP}$ and $G_{\rm RP}$ bands based on its 36 months' observation. LAMOST database has accumulated more than 8 million stellar spectra in DR5 with a spectral resolution of R $\sim$ 1800. Effective temperature $T_{\rm eff}$, surface gravity log $g$, and metallicity [Fe/H] are delivered from the spectra through the LAMOST Stellar Parameter pipeline (LASP; \citealp{wu2011}) with the precision of about 110 K, 0.2 dex, and 0.1 dex, respectively \citep{luo2015}. \citet{xiangdd} use a data-driven Payne approach to provide abundances of 6 million stars for 16 elements as well as the [$\alpha$/Fe] with the precision of about 0.03 dex, which is defined as a weighted mean of [Mg/Fe], [Si/Fe], [Ca/Fe], and [Ti/Fe]. The results are publicly available as a value-added catalog of the LAMOST DR5.
Note that the stellar parameters obtained by LAMOST assume that the spectrum only has contribution from one star. In 
case of binaries, the stellar parameters may suffer systematic errors. Fortunately, 
the typical resulting systematic errors are negligible for LAMOST-like spectra, 0.05 dex for [Fe/H] and [$\alpha$/Fe], 100 K for $T_{\rm eff}$ within our temperature ranges \citep{bi_specfit}.

To grantee the quality of the sample, the following constraints are required:

For \textit{Gaia} data:
\begin{itemize}
   \item [1)]
   \texttt{phot}$\_$\texttt{bp}$\_$\texttt{mean}$\_$\texttt{flux}$\_$\texttt{over}$\_$\texttt{error} $>$ 100
\item [2)]
\texttt{phot}$\_$\texttt{g}$\_$\texttt{mean}$\_$\texttt{flux}$\_$\texttt{over}$\_$\texttt{error} $>$ 100
  \item [3)]
\texttt{phot}$\_$\texttt{rp}$\_$\texttt{mean}$\_$\texttt{flux}$\_$\texttt{over}$\_$\texttt{error} $>$ 100
\item [4)]
\texttt{phot}$\_$\texttt{proc}$\_$\texttt{mode} = 0
\item [5)]
\texttt{duplicated}$\_$\texttt{source} = \texttt{False}
\end{itemize} 

$\quad$For LAMOST data:
\begin{itemize}
   \item [6)]
the signal-to-noise ratios for the $g$ band (S/N$_g$) are larger than 20
    \item [7)]
$T_{\rm eff} >$ 4500 K 
\end{itemize}

$\quad$For spatial location:

\begin{itemize}
   \item [8)]
$E(B-V)$ $<$ 0.05 mag according to the \citeauthor*{sfd} (\citeyear{sfd}, hereafter SFD) dust reddening map
    \item [9)]
Galactic latitude $|b| > 20$ deg
\item [10)]
 vertical distance to the Galactic disk $|Z| > 0.2$ kpc
\end{itemize} 

\begin{figure}[htbp] 
    \centering 
    \includegraphics[width=3.5in]{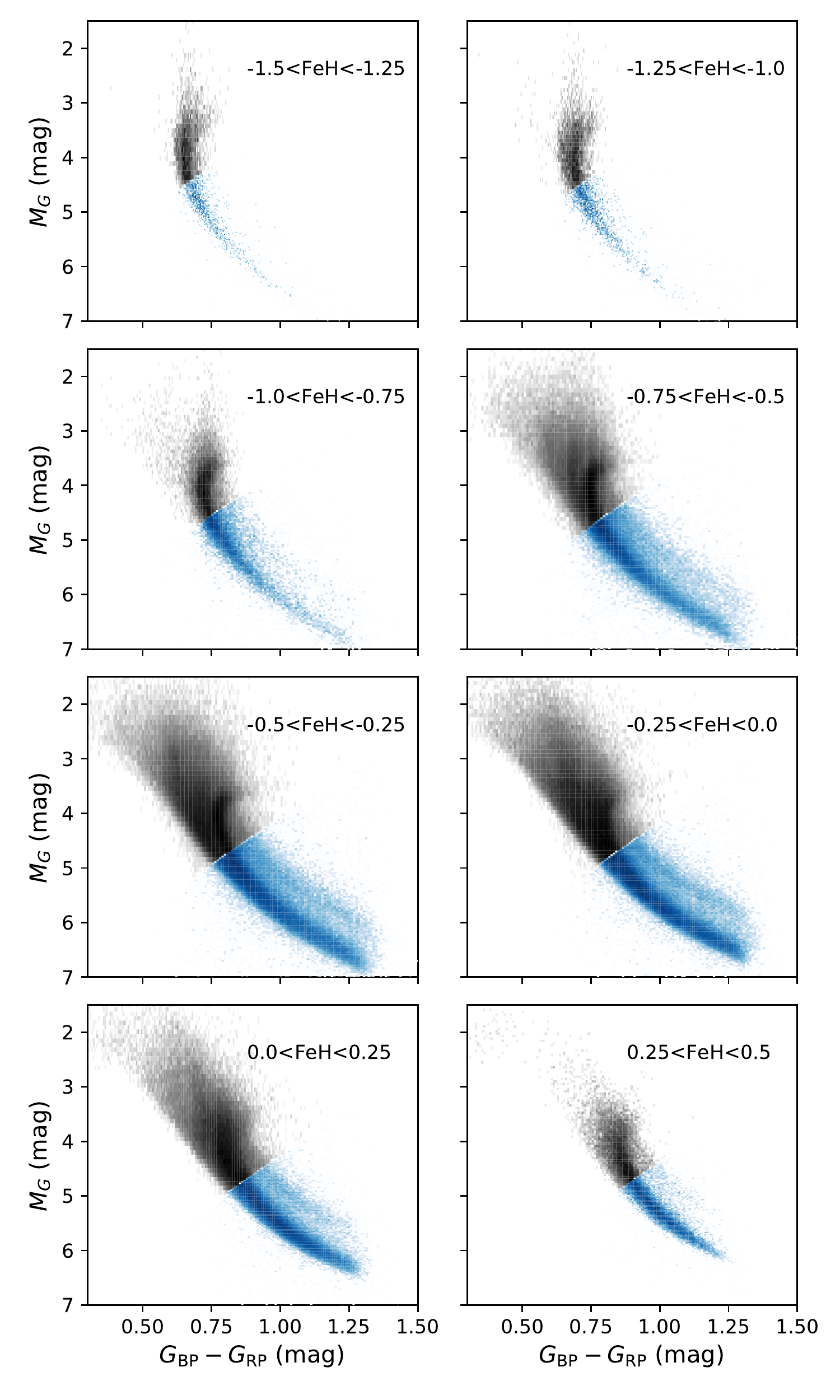}  
    \caption{Selections of the MS stars for different metallicity bins. Blue points are selected. Metallicities are labelled in each panel.\label{turnoff}} 
\end{figure}

\begin{figure}[htbp] 
    \centering 
    \includegraphics[width=3.5in]{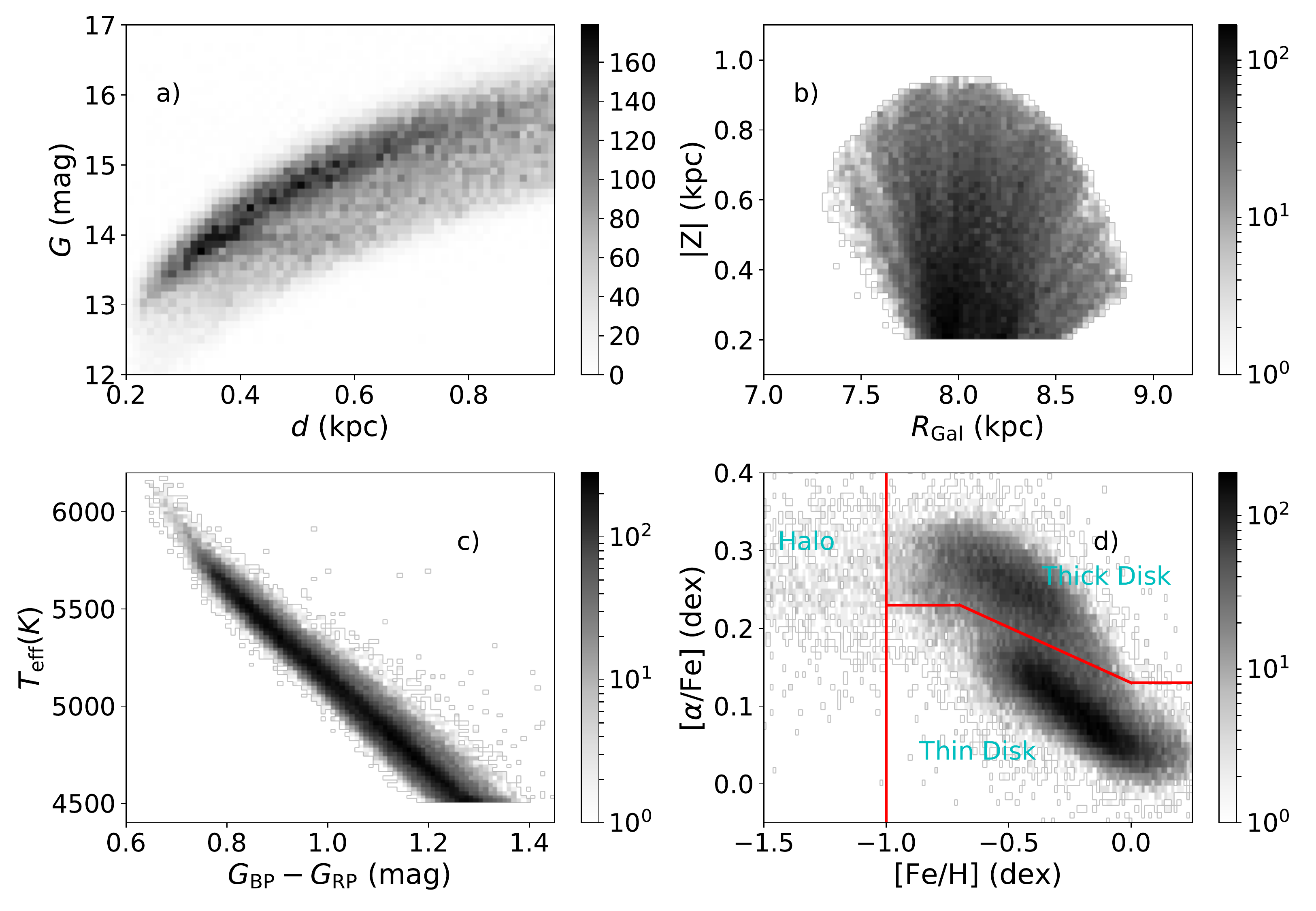}  
    \caption{Stellar parameters distributions of the sample in this paper. Panel $a$: distance $d$ against the $G$ magnitude. Panel $b$: Galactocentric distances $R_{\rm Gal}$ against vertical distance from the disk $|Z|$. Panel $c$: intrinsic $G_{\rm BP}-G_{\rm RP}$ against effective temperature $T_{\rm eff}$. Panel $d$: [$\alpha$/Fe] against [Fe/H]. Color bars indicate number densities. Halo, thin disk, and thick disk stars are distinguished by an empirical cut. \label{data}} 
\end{figure}

The corrected $G$ magnitudes \citep{gaia2020,phot_cont} of the sources with 6-parameter astrometric solutions are used in this paper. 
Meanwhile, previous works have revealed that \textit{Gaia} photometry suffers magnitude-dependent systematic errors up to 10 mmag \citep{maw2018,casa2018}. 
Further color and magnitude corrections are applied.
The former is from \citet{niudr2,niudr3}, which present precise corrections on colors using the spectroscopy-based stellar color regression method \citep{2015scr}, achieving a precision of about 1 mmag.  
The latter is provided by \citet{yl}, where a machine learning technique is applied to train the observed $UBVRI$ magnitudes of about 10,000 Landolt standard stars into the \textit{Gaia} EDR3 magnitudes. 
For multiply observed \textit{Gaia} objects in LAMOST, we combine their epoch $T_{\rm eff}$, log $g$, [Fe/H] and [$\alpha$/Fe] values using inverse variance weighting.

Because we have specified the spatial location, 
we assume that all sources are beyond the source of reddening. 
All colors referred to hereafter are dereddened with the SFD dust map and the empirically determined reddening coefficients $R(G-G_{\rm RP})$ and $R(G_{\rm BP}-G_{\rm RP})$, which are dependent on temperature and reddening, as provided in \citet{niudr3}. 
We have tested other 2D dust maps like \citet{lenz2017} and \citet{planck} for reddening correction, and found no big differences.
The absolute \textit{Gaia} magnitude in the $G$ band is calculated by $M_G=G+5+5{\rm log_{10}}(\varpi/1000)-A_G$, where $\varpi$ is the \textit{Gaia} parallax and $A_G=1.890$ $E(G_{\rm BP}-G_{\rm RP})$ according to \citet{ws}.

We select the main sequence (MS) stars in the $T_{\rm eff}$ and log $g$ diagram using the same criterion in \citet{niudr2}. As shown in Figure \ref{turnoff}, main sequence turn-off (MSTO) stars are eliminated in the Hertzsprung-Russell diagram following the criteria in Appendix A, as their intrinsic magnitudes can not be precisely determined by their colors and metallicities alone.

To avoid Malmquist bias, i.e., multiple systems can be detected further away than single stars in a magnitude-limited sample, distances are limited to 0.95 kpc. Therefore the upper limit of the $G$ mag is about 16.5 where the color and magnitude corrections are valid.

Since one of the main purpose in this paper is to investigate the relationship between the $\alpha$ abundances and the binary fractions, we limit our sample to be within a range from 0.55 to 0.8 solar mass. A wide $\alpha$ abundances range is engaged by this.
Stellar mass is estimated by interpolating the color mass relation provided by the PARSEC (\citealp{parsecbre,parsectang}) model at every 0.1 dex from $-$1.5 to 0.5 dex of [Fe/H], as shown in Figure \ref{cm} (see Appendix B).

Finally, our sample contains 86,141 objects. The distributions of stellar parameters of our sample are shown in Figure \ref{data}. 
In panel $a$, the discontinuity of the magnitude distribution around $G \approx 14$ mag is due to the selection functions of the LAMOST \citep{niudr2} and has no impact on this work. 
Panel $b$ shows the spatial distributions. Panel $c$ gives the distributions of the intrinsic color and effective temperature. Our sample covers wide ranges of [Fe/H] of about 2 dex and [$\alpha$/Fe] of 0.5 dex. 
As expected, a bimodal distribution of [$\alpha$/Fe] is clearly seen in panel $d$, corresponding to the chemically defined Galactic thin and thick disks \citep{Fuhrmann1998,Fuhrmann2004}. By applying an empirical cut plotted as red lines in the [Fe/H]-[$\alpha$/Fe] plane, thin disk, thick disk, and halo stars are grouped. Some contamination may be found in the three populations. But it should not affect the main results of this work, considering that the typical precision of [Fe/H] and [$\alpha$/Fe] of our sample stars are 0.1 and 0.03 dex, respectively.

\section{Method} \label{sec:method}

\subsection{SLOT method overview}

Proposed by \citet{2015loci2}, the SLOT method is developed to provide model-free estimations of binary factions for large numbers of stars. They have applied the SLOT method to two samples of Stripe 82 stars by combing the recalibrated SDSS photometric data with the spectroscopic information from the SDSS and LAMOST surveys. After fitting the $u-g$, $g-r$, $r-i$, and $i-z$ colors as a function of the $g-i$ color and [Fe/H], they find that the fitting residuals are asymmetric, pointing to the presence of a significant population of binaries.
This method works well for not only close binaries but unresolved wide binaries as well. It is also insensitive to the assumed mass ratio distribution and does not require multiple epoch data.

\begin{figure}[htbp] 
    \centering 
    \includegraphics[width=3.5in]{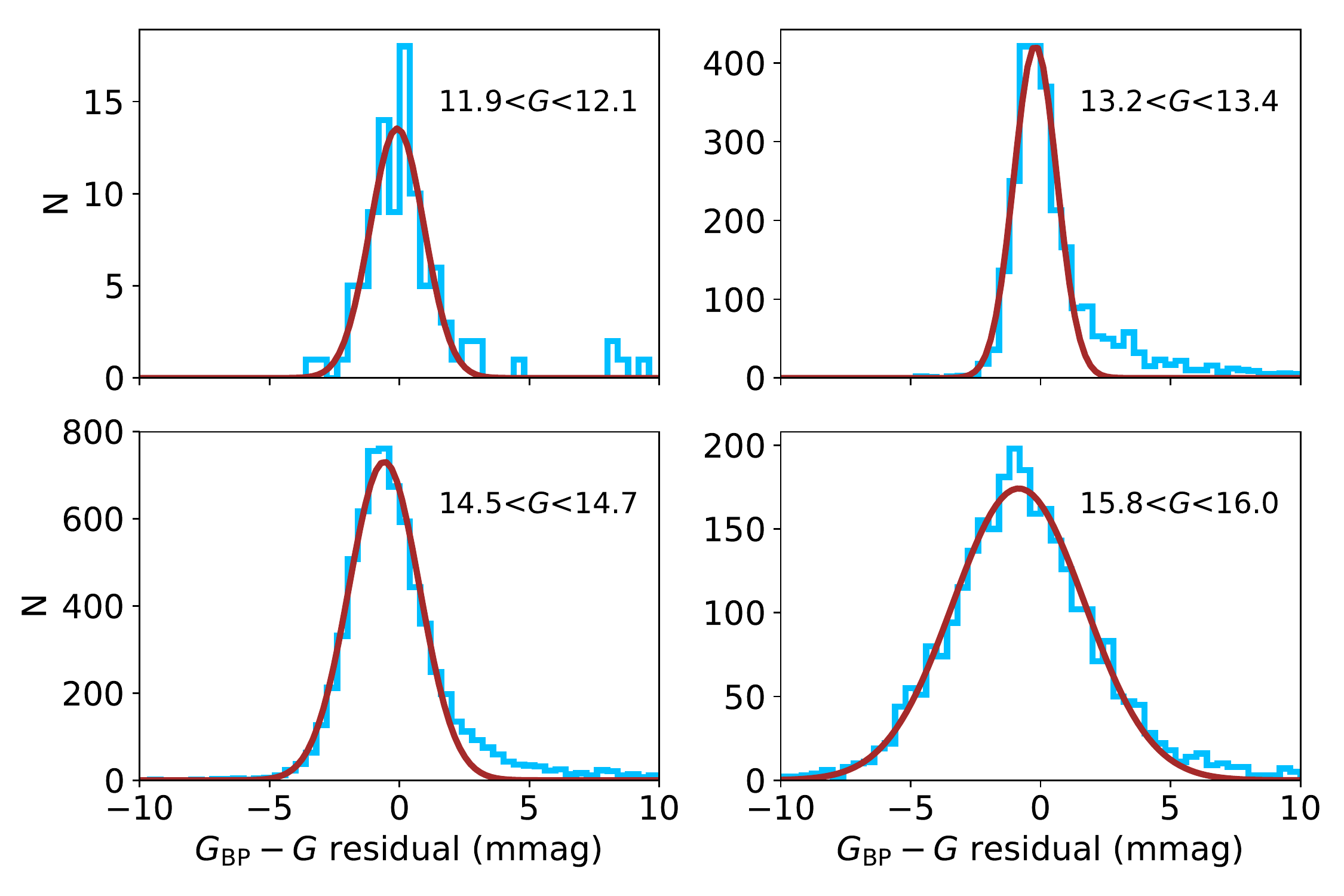}
    \caption{\textit{blue}: Distributions of the $G_{\rm BP}-G$ residuals in the different $G$ magnitude ranges. \textit{red}: Results of the Gaussian model fitting.\label{ex}} 
\end{figure}

In \citet{niudr3}, we have fitted the metallicity-dependent stellar loci for the \textit{Gaia} colors (see Table \ref{tab1}) and found the asymmetric $G_{\rm BP}-G$ residuals (defined as loci predicting colors $-$ observed colors, hereafter the same) after correcting the magnitude-dependent systematic effects of colors. To be more clearly, Figure \ref{ex} shows the distributions of the $G_{\rm BP}-G$ residuals at different $G$ magnitudes from 12 to 16 mag. The observed $G_{\rm BP}-G$ residuals, as shown in the blue histograms, are fitted with Gaussian models. The results are over-plotted in the red lines. There are significant discrepancies between the symmetric Gaussian distributions (red) and the observed $G_{\rm BP}-G$ residuals (blue) in the first 3 panels. The discrepancies weakens in the last panel when $G \sim 15.9$ mag, due to the larger photometric errors at fainter magnitudes\footnote{The SLOT method is less valid when photometric errors are much larger than the maximum locus offset caused by binary stars. The detailed number depends on the colors used, 3 mmag in the case of this work (see Figure 6 in the next subsection).}.
Fortunately, almost all the sample stars in our study are brighter than 16.5 mag in the $G$ band, thus they do not suffer from this problem.

\begin{figure}[htbp] 
    \centering 
    \includegraphics[width=3.5in]{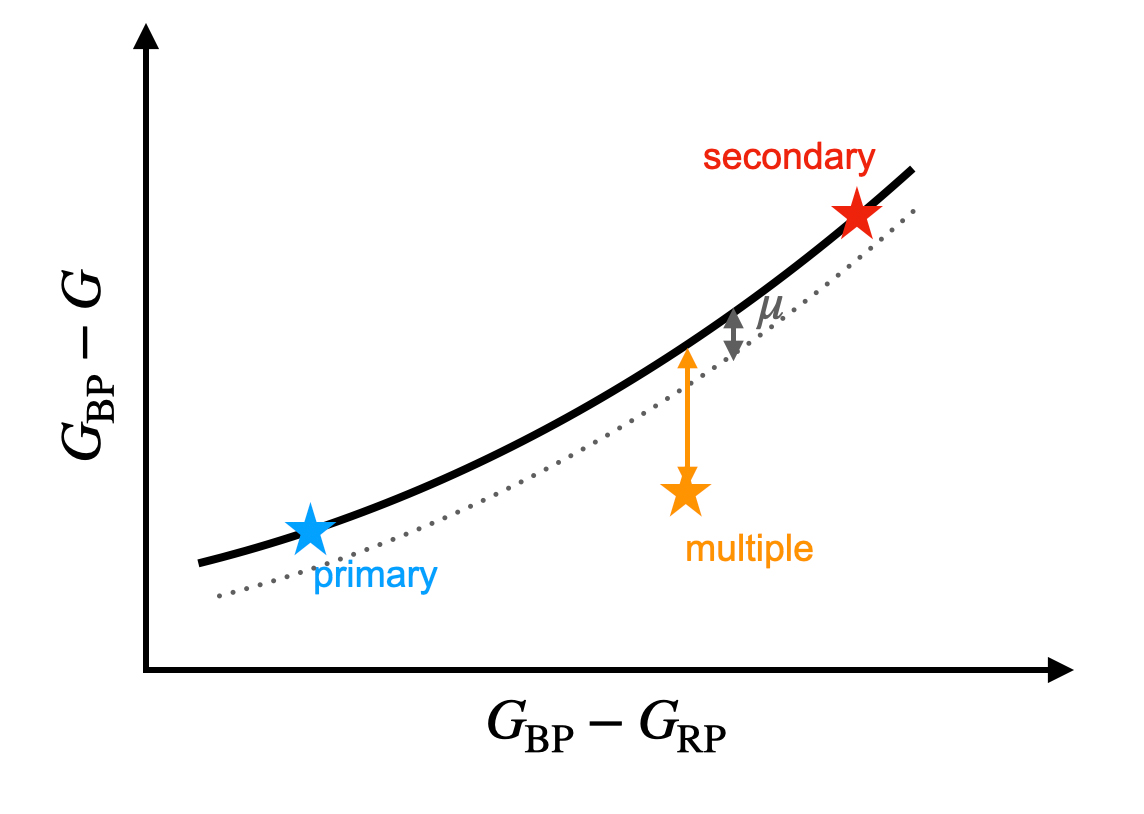}
    \caption{Overview of the SLOT method. Black solid line donates the stellar locus of MS single stars of a given metallicity, where places the primary and the secondary stars. The combination of single stars lies below the stellar locus. Gray dotted line represents the empirically measured stellar locus, with an systematic offset $\mu$ labelled. Scale of the binary effects is plotted exaggeratedly for clarity. \label{ex0}} 
\end{figure}

As explained by Figure\,1 of \citet{2015loci2}, the SLOT method estimates the binary fraction of a given sample by modeling the distributions of the observed color residuals with respect to the metallicty-dependent stellar loci. Here we reproduce the illustration as Figure \ref{ex0} for better reading. For a given metallicity, single stars lie on the stellar locus, while binary/multiple systems deviate systematically from the locus. 
Two sets of Monte Carlo simulations are performed, assuming that all stars in the sample are either single or binary.
Then by adjusting the relative fraction of stars in the two sets of simulations to fit the observed residual distributions, the binary fraction of the sample is determined. 
On the other hand, due to the existence of binary stars in the sample when fitting the metallicity-dependent stellar loci, systematic offsets $\mu$ are involved (shown by the gray dotted line in Figure \ref{ex0}). 
It only causes the mean values of the distributions of the observed $G_{\rm BP}-G$ residuals not strictly zero, shifting less than 1 mmag from Figure \ref{ex}, while the mean values of the simulated distributions are zero. 
Note that the $\mu$ and binary fraction $f_{\rm b}$ are tightly correlated. For convenience, both the $\mu$ and $f_{\rm b}$ are set as free variables in this work. In Section 4, one can see that the results of $\mu$ and $f_{\rm b}$ are strongly correlated as expected.

A minimum $\chi^2$ technique is adopted in the fitting: 

\begin{gather}\label{eqchi}
    \chi^2=\sum_{i=1}^{N_{\rm bin}}\frac{(N_{\rm obs}^i-N_{\rm sim}^i)^2}{{N_{\rm obs}^i} \times (N_{\rm bin}-1)} 
\end{gather}

\begin{gather}\label{simcom}
     N_{\rm sim}^i=N_{\rm binary}^i \times f_{\rm b} + N_{\rm single}^i \times (1-f_{\rm b})
\end{gather}

Considering the maximum of the binary influences on the $G_{\rm BP}-G$ residuals and various uncertainties (see the next subsection), when calculating the $\chi^2$, $G_{\rm BP}-G$ residuals are equally binned in the range from $-$7 to 4 mmag with a bin size of 0.275 mmag, because the simulated residuals barely reach values beyond these ranges. 
$N_{\rm obs}^i$, $N_{\rm single}^i$, and $N_{\rm binary}^i$ are the star numbers that are observed, predicted by the single star simulation, and predicted by the binary star simulation in the $i$th bin, respectively.
Both the single and binary sets of simulations are carried out 100 times to eliminate the random errors.
For a given binary fraction $f_{\rm b}$ and an offset $\mu$, the residuals of the single and binary sets are combined according to Equation \ref{simcom}, and then shifted toward the bluer side by the offset $\mu$.
$N_{\rm sim}^i$ is the finally predicted star number by the simulation in the $i$th bin.
$N_{\rm bin} = 40$ is the bin number of the histogram. 
We vary $f_{\rm b}$ from 0.0 to 1.0 at 0.01 intervals and $\mu$ from 0.0 to 1.5 mmag at 0.01 mmag intervals. $\chi^2$ is calculated for each set of $f_{\rm b}$ and $\mu$. By choosing the global minimum $\chi^2$, the best solution is determined. 

In order to explore the binary fraction of the field star as a function of metallicities, we divide the whole sample into 35 bins in the [Fe/H] and  [$\alpha$/Fe] diagram. Bin width for [Fe/H] is fixed to be 0.1 dex, and it is flexible for [$\alpha$/Fe] depending on the number density. 

We use the Bootstrap Sample method to evaluate formal errors of the binary fractions. 
Random sampling is done by 1,500 times for one bin containing the most objects of 4,801.
We fit the distribution of 1,500 $f_b$ values by a Gaussian model and get $\sigma$=0.04 as the formal error of the binary fraction of that bin. 
Errors for other bins are then estimated assuming they are inversely proportional to the square of the ratio of their numbers to 4,801,
e.g., about 0.1 for a sample size of 1000.    

\begin{table}[H]
\footnotesize
\centering
    \caption{Fitting Coefficients of the Metallicity-dependent Stellar Color Loci\label{tab1}}
    \renewcommand\tabcolsep{3.0pt}
    \begin{tabular}{|c|cc|c|cc}

Coeff.$^1$ & dwarf$^2$ & RGB$^3$  & Coeff. & dwarf &  RGB \\
\hline
$a_0$ & 6.01E-03  & $-$1.41E-02  & $a_8$ & $-$3.08E-03  & 6.63E-04  \\
$a_1$ & $-$3.22E-03& 1.34E-02  & $a_9$ & 1.62E-01  & 1.51E-02  \\
$a_2$ &  5.98E-03 & $-$4.10E-03  & $a_{10}$ &  $-$1.63E-02 &  2.45E-02 \\
$a_3$ &  3.92E-03 & $-$5.19E-04  & $a_{11}$ & 8.05E-04  &  $-$1.81E-03 \\
$a_4$ &  5.70E-04 & $-$6.29E-05  & $a_{12}$ & $-$3.68E-02  & 5.72E-02  \\
$a_5$ &  2.82E-01 & 3.75E-01  & $a_{13}$ & 3.15E-03  & $-$8.66E-03  \\
$a_6$ &  2.21E-02 & $-$2.33E-02  & $a_{14}$ & 5.28E-03  &  $-$1.54E-02 \\
$a_7$ &  $-$5.86E-03 & 7.35E-03  \\
 \end{tabular}
 \begin{tablenotes}
     \item [1]$^1$ $G_{\rm BP}-G$=$f(x,y)=a_0+a_1\times y+a_2\times y^2+a_3\times y^3+a_4\times y^4+a_5\times x+a_6\times x\times y+a_7\times x\times y^2+a_8\times x\times y^3+a_9\times x^2+a_{10}\times x^2\times y+a_{11}\times x^2\times y^2+a_{12}\times x^3+a_{13}\times x^3\times y+a_{14}\times x^4$, where $x$ is $G_{\rm BP}-G_{\rm RP}$ and $y$ is [Fe/H].
     \item [2]$^2$ Coefficients are fitted under the $G_{\rm BP}-G_{\rm RP}$ range from 0.6 to 1.4 mag and [Fe/H] range from $-$1.5 to 0.5 dex. 
     \item [2]$^3$ Coefficients are fitted under the $G_{\rm BP}-G_{\rm RP}$ range from 0.75 to 1.7 mag and [Fe/H] range from $-$1.5 to 0.5 dex. 
\end{tablenotes}
\end{table}

\subsection{Descriptions of the simulation}

\begin{figure*}[htbp]
    \centering
    \includegraphics[width=7in]{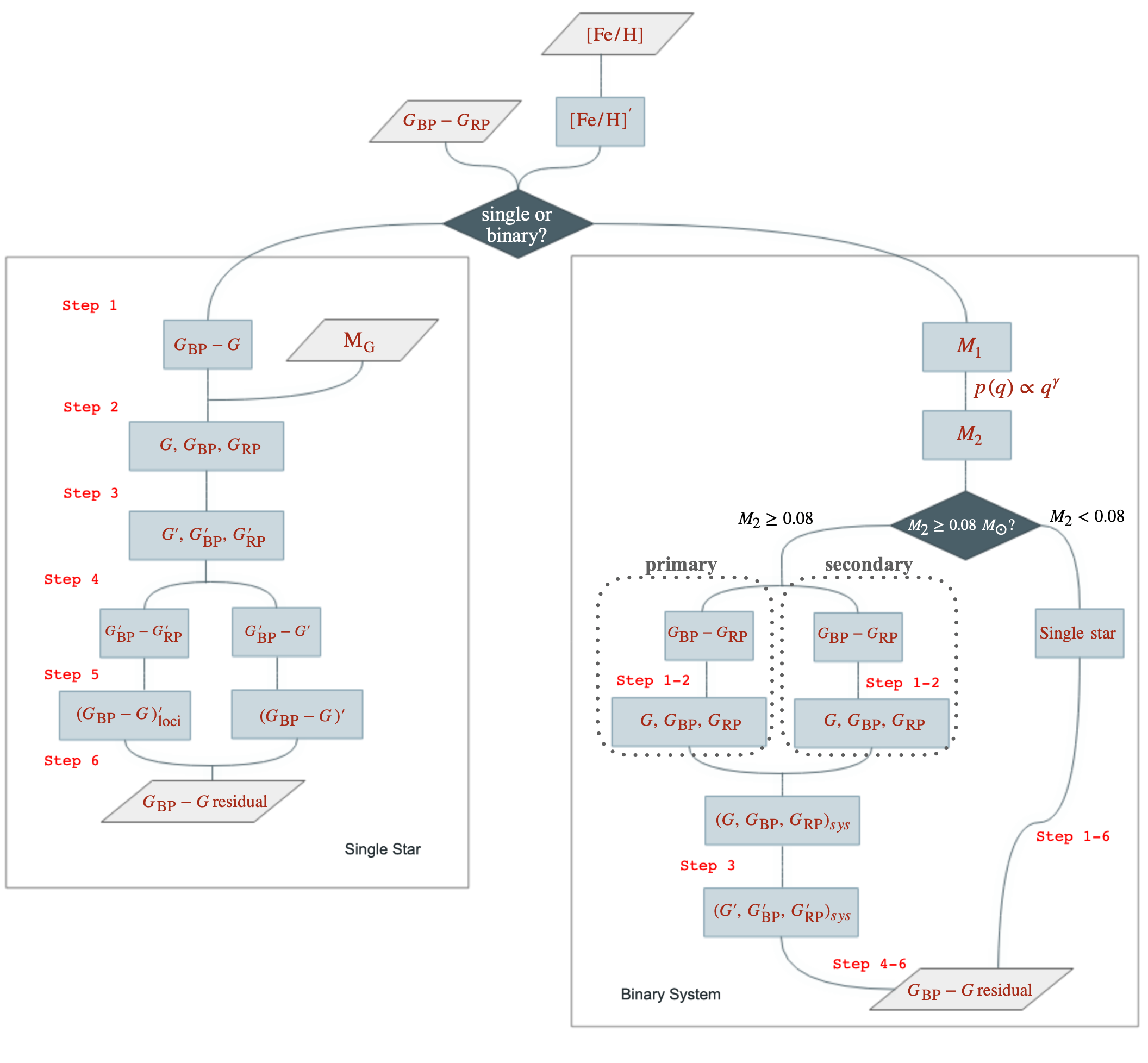}
    \caption{Schematic illustration of the SLOT method in \textit{Gaia} photometry. The observed [Fe/H], absolute $G$ magnitude, and $G_{\rm BP}-G_{\rm RP}$ are adopted as inputs. $G_{\rm BP}-G$ residuals of the single set and binary set are the output. Each step is described in the main text in details.}
    \label{flow}
\end{figure*}

Here we describe simulations of the binary and single sets step by step. A schematic description is shown in Figure \ref{flow}. For each object, the observed $G_{\rm BP}-G_{\rm RP}$, [Fe/H], and absolute $G$ magnitude are the inputs. Error in this work is assigned and added through a normal distribution. We use superscript $'$ to indicate values after adding errors. For example, [Fe/H]$'$ is sampled by adding the random error $\sigma_{\rm ran}$([Fe/H]) and systematic error $\sigma_{\rm sys}$([Fe/H]) to the [Fe/H].

When simulating the single set, steps 1--6 are as below:

\textit{Step 1}:
Given the intrinsic $G_{\rm BP}-G_{\rm RP}$ and [Fe/H]$'$, the intrinsic $G_{\rm BP}-G$ is calculated with the metallicity-dependent loci in Table \ref{tab1}.

\textit{Step 2}:
The absolute $G_{\rm BP}$ and $G_{\rm RP}$ magnitudes are obtained by combining the absolute $G$ magnitude and the intrinsic colors.

\textit{Step 3}:
Random errors in the three magnitudes are added.

\textit{Step 4}:
Then $G_{\rm BP}'-G'$ and $G_{\rm BP}'-G_{\rm RP}'$ are derived.

\textit{Step 5}:
When yielding the simulated $(G_{\rm BP}-G)'$, $\sigma_{\rm ZP ,sys}$ (see the next subsection) is acquired according to the $G$ magnitude and added to the $G_{\rm BP}'-G'$. $(G_{\rm BP}-G)_{\rm loci}'$ is computed using $G_{\rm BP}'-G_{\rm RP}'$ and [Fe/H]$'$ by the loci at the same time.

\textit{Step 6}:
Simulated $G_{\rm BP}-G$ residual defined as the difference between $(G_{\rm BP}-G)_{\rm loci}'$ and $(G_{\rm BP}-G)'$ is the output.

When simulating the binary set, we assume that the observed $G_{\rm BP}-G_{\rm RP}$ is belong to the primary as it is the brighter component. Then the primary mass is determined based on the PARSEC model. The binary mass ratio distribution is assumed to follow a power law with the index $\gamma$= 0.3 \citep{2013araa}. 
Mass of the secondary is obtained with a mass ratio $q$. 
When $M_2 < 0.08 M_{\odot}$, 
we treat the binary system as one single star, performing the simulation following steps 1--6. When $M_2 \geq 0.08 M_{\odot}$, the intrinsic $G_{\rm BP}-G_{\rm RP}$ of the secondary is interpolated based on the same PARSEC model assuming the metallicity of the binary system is uniform. 
Next, the absolute magnitudes of the primary and the secondary in three bands are separately calculated as steps 1 and 2 do, except the absolute $G$ magnitudes are interpolated from the PARSEC model (see Appendix). Then the combined absolute magnitudes of the binary system are derived. Finally steps 3--6 are applied to the combined magnitudes and colors, $G_{\rm BP}-G$ residual of the binary system are yielded.

\begin{figure}[htbp] 
    \centering 
    \includegraphics[width=3.5in]{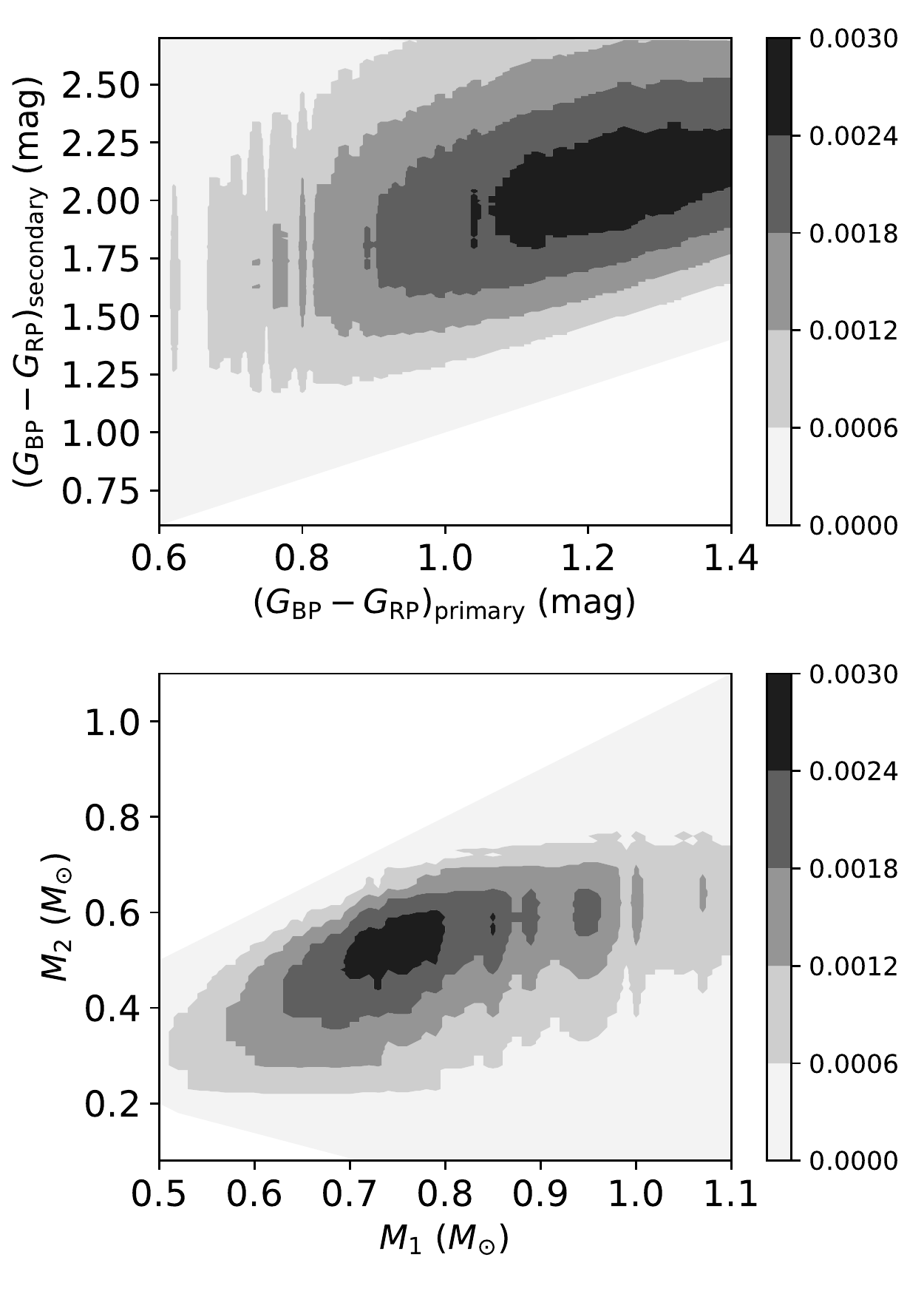}  
    \caption{Distributions of differences between the combined $G_{\rm BP}-G$ colors of binary systems and that predicted by the metallicity-dependent stellar loci for the combined $G_{\rm BP}-G_{\rm RP}$ color of the system. Binary systems are assumed to be composed of two single dwarf stars with the solar abundance.\label{cc}} 
\end{figure}

Following the above flow chart, we could estimate the $G_{\rm BP}-G$ residual for binary systems of different combinations of $G_{\rm BP}-G_{\rm RP}$ and mass. In Figure \ref{cc}, the residuals are estimated assuming all errors equal zero and [Fe/H]=0. We can see that the maximum binary influence on the $G_{\rm BP}-G$ residual is about 3 mmag, happening at the intermediate mass ratios. That helps to set the boundaries when calculating the $\chi^2$.

\subsection{Error treatments}\label{err}

As described in the previous subsection, different kinds of errors are considered in the simulation, including random and systematic errors of LAMOST [Fe/H] and \textit{Gaia} photometry.

The random errors $\sigma_{\rm ran}$([Fe/H]) are fitted using multi-observed objects as a function of S/N$_g$ and [Fe/H] with a third-order polynomial. We only take observations with S/N$_g < 100$ in the fitting to avoid overfitting. When assigning $\sigma_{\rm ran}$([Fe/H]) for stars with S/N$_g > 100$, S/N$_g = 100$ are used. For S/N$_g$ = 20, the $\sigma_{\rm ran}$([Fe/H]) are about 0.045, 0.034, and 0.027 dex for [Fe/H] = $-$1, $-$0.5, and 0 dex, respectively. For S/N$_g$ = 50, the $\sigma_{\rm ran}$([Fe/H]) are about 0.029, 0.022, and 0.018 dex for [Fe/H] = $-$1, $-$0.5, and 0 dex, respectively. Systematic errors $\sigma_{\rm sys}$([Fe/H])$= 0.05 - 0.05 \times$ [Fe/H] \citep{2015loci2} are also considered.

For each object, \textit{Gaia} EDR3 catalog provides flux errors of three \textit{Gaia} bands. They are converted into magnitude errors as follows:

\begin{gather}
    m = -2.5 \times \log_{10}(F) + ZP \notag \\
    m + \sigma_{m} = -2.5 \times \log_{10}(F - \sigma_{F}) + ZP \\
    \sigma_{m} = m + \sigma_{m} - m \notag
\end{gather}

where $F$ and $\sigma_{F}$ are respectively the published flux and flux error, $ZP$ is the photometric zero point, $\sigma_{m}$ is the magnitude error.

The uncertainties of the photometric zero points (ZP) in the VEGAMAG, $\sigma_{\rm ZP}$ = 2.8 mmag for all three bands, are provided in \citet{phot_cont}. However, the uncertainties in the three bands are probably strongly correlated. Therefore, the uncertainties in terms of colors should be much smaller than involving $\sigma_{\rm ZP}$ of two bands independently. 
Besides, $\sigma_{\rm ZP}$ are very likely to be magnitude-dependent, considering the unique observation mode of \textit{Gaia} and the magnitude-dependent systematic effects discovered previously \citep{niudr2,maw2018,casa2018}.

We are mostly concerned with the effect of $\sigma_{\rm ZP}$ on the $G_{\rm BP}-G$ residuals in this work. 
To get an empirical estimation of this effect, which is represented by $\sigma_{\rm ZP,sys}$ in this work, red giant branch (RGB) stars are used as ideal single stars in terms of photometry. Because in RGB-MS systems, the RGB stars are much brighter than the MS stars. 
A study of double-lined spectroscopic binaries (SB2) in APOGEE DR16 and DR17 \citep{sb2apg} shows that only a few percent of SB2 appear to be RGB-RGB system. Even in that case, the asymmetric binary effect is tiny since the colors/masses of the two components should be similar.
Therefore, $G_{\rm BP}-G$ residuals of RGB stars should be fully accounted by errors of [Fe/H], random errors of the three bands, and the $\sigma_{\rm ZP,sys}$.

\begin{figure}[htbp] 
    \centering 
    \includegraphics[width=3.2in]{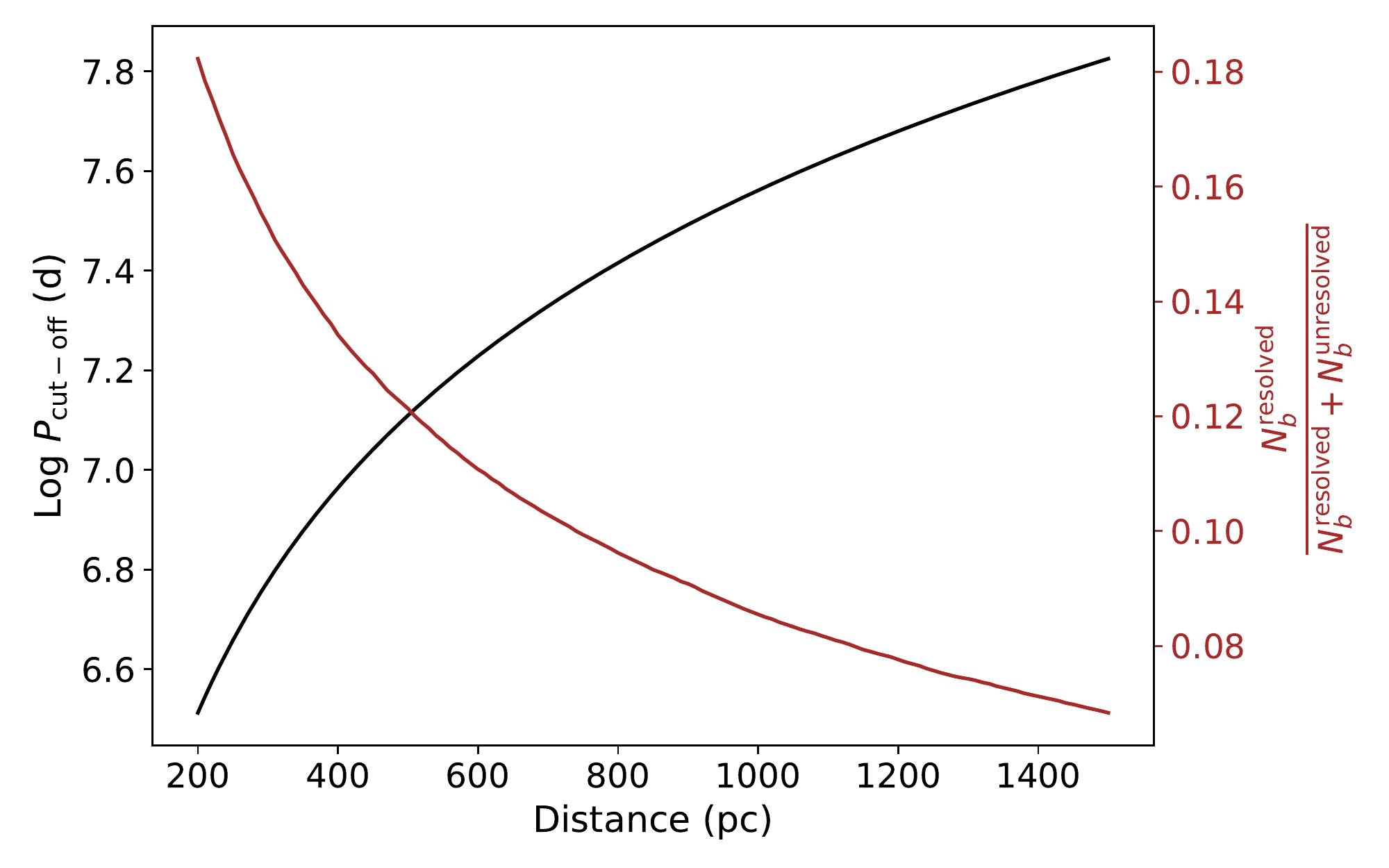}
    \caption{\textit{black line}: Logarithmic cut-off period as a function of distance with a total mass of 1 M$_\odot$. \textit{red line}: Fractions of missing resolved binaries. Normal distributed phases and inclination angles of the binary system is adopted. The median value is chosen at each distance. 
\label{fmiss}} 
\end{figure}

\begin{figure}[htbp] 
    \centering 
    \includegraphics[width=3in]{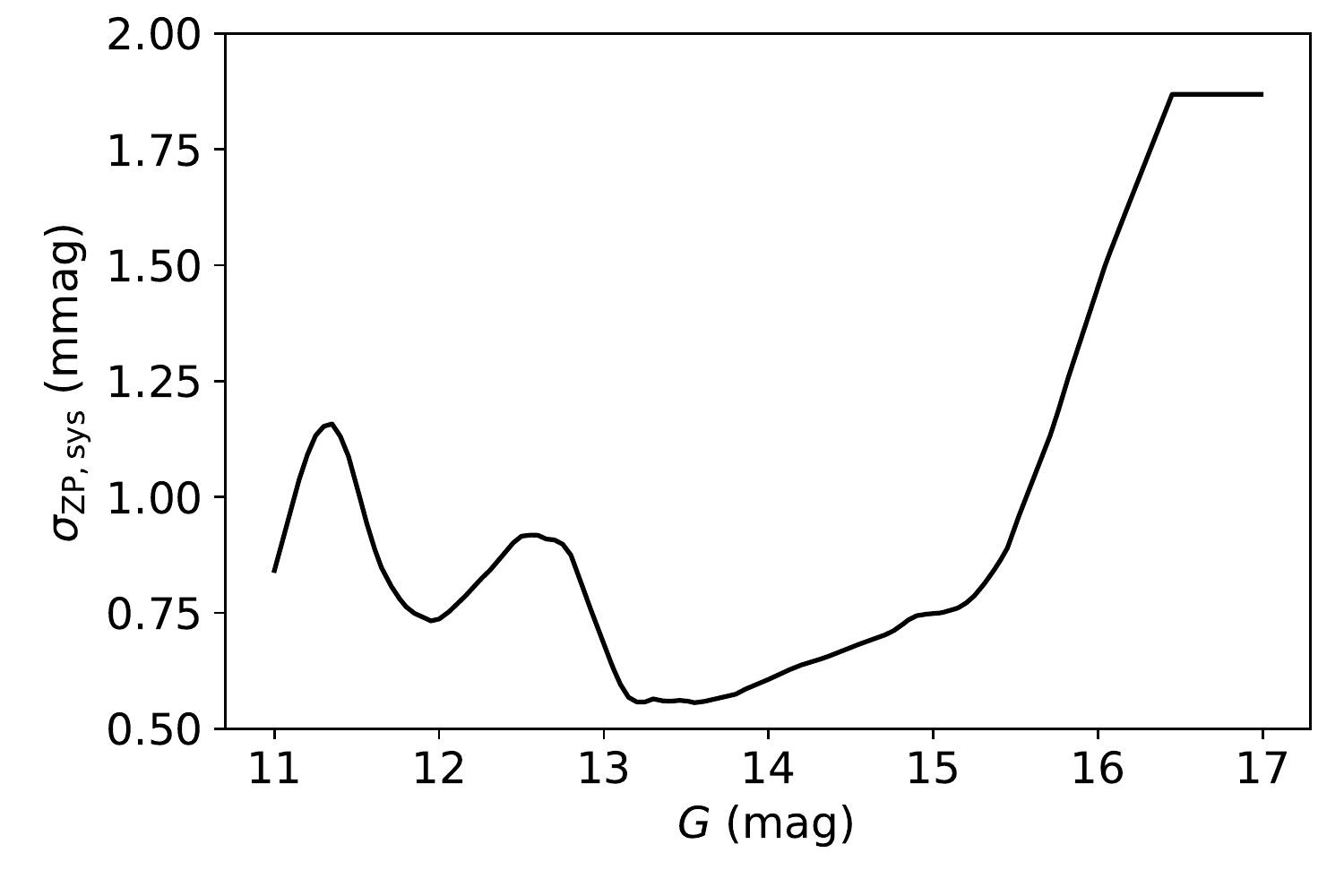}  
    \caption{Substitutions of the uncertainties of the $ZP$ derived from the RGB stars against the $G$ magnitude. \label{fsys}} 
\end{figure}

A number of 56,871 RGB stars are selected in the same way as \citet{niudr2}. Corrections of the \textit{Gaia} magnitudes and colors mentioned in Section \ref{sec:data} are also performed. The matallicity-dependent stellar loci of the RGB stars are listed in Table \ref{tab1} as well. 
In order to explore whether $\sigma_{\rm ZP,sys}$ is magnitude-dependent, the RGB stars are divided into bins of 0.2 mag width in the range from 11 to 16.4 mag. 
For each bin, we use the single star model shown in Figure \ref{flow} to fit their $G_{\rm BP}-G$ residuals. The $\sigma_{\rm ZP,sys}$ is varied from 0.0 to 2 mmag at 0.01 mmag interval, the corresponding $\chi^2$ values are calculated. 
By choosing the global minimum $\chi^2$, the best solutions of the $\sigma_{\rm ZP,sys}$ against the $G$ magnitudes are determined. 
Figure \ref{fsys} plots the smoothed curve of the $\sigma_{\rm ZP,sys}$. We can see that $\sigma_{\rm ZP,sys}$ is magnitude-dependent, showing the minimum at around $G \sim 13.5$ mag where owns the best data quality \citep{phot_cont}. It varies around 0.75 mmag when $G < 15.5$ mag, and then increases quickly towards fainter magnitudes. Only a few RGB stars (dwarfs as well) are of $G > 16.4$ mag, therefore the value of $\sigma_{\rm ZP,sys}$ at $G = 16.4$ mag is adopted for stars with $G > 16.4$ mag.

\subsection{Correction for resolved wide binaries}

In the above simulation, we have assumed that binary stars are unresolved. However, due to the angular resolution of telescopes, 2$^{\prime\prime}$ for \textit{Gaia} \citep{2018Arenou} and 2.5$^{\prime\prime}$ for LAMOST \citep{liuchao2019}, a binary star would be resolved and mistakenly identified as two single stars when its two companions have a recognizable spatial separation $\rho > 2^{\prime\prime}$. Given the total mass of the binary system and the separation between two components, the cut-off period $P_{\rm cut-off}$ could be estimated by the Kepler's Laws. Binary fractions with periods larger than the $P_{\rm cut-off}$ would be underestimated. We correct for the effect of the resolved binaries in this subsection.

According to the target selection process of LAMOST \citep{yuan2015lamost}, targets are required to have no neighbour within 5$^{\prime\prime}$ radius that is brighter than (m+1) mag, where m is magnitude of a given target star. 
Therefore, the fainter secondary has a little chance of being targeted in the LAMOST survey. We assume that LAMOST only targets the primary stars of the resolved binaries.
We use $N_s$, $N_b^{\rm unresolved}$, and $N_b^{\rm resolved}$ to represent the number of single stars, unresolved binary systems, and primaries of the resolved binary systems in our sample, respectively. The binary fractions in the above simulation can be described as: 
\begin{equation} \label{fb}
    f_b=\frac{N_b^{\rm unresolved}}{N_b^{\rm unresolved}+N_b^{\rm resolved}+N_s}
\end{equation}

Given the orbit distribution of the binary systems, the fractions of the resolved ones could be estimated via Monte Carlo simulations. As suggested by \citet{raghavan2010}, we choose the period distribution to follow a log-normal Gaussian profile with a mean of log$P$ = 5.03 and $\sigma_{{\rm log}P}$ = 2.28, where $P$ is in unit of day. 
A total mass of 1 M$_\odot$ is also assumed. 
We have verified that the result changes slightly with the assumed total mass. 
We further adopt uniform distributions of orbital inclinations and phases. 
In addition, studies have shown that the average eccentricity increases with the binary period but in a complex way \citep{2003Halbwachs,2016Tokovinin,moe2017}. The resolved binaries only contribute a few percent, not the main concern in this paper. Hence circular orbits are applied.
The cut-off periods at different distances is firstly calculated and plotted in black in Figure \ref{fmiss}.
Based on the above assumptions and the angular resolution of \textit{Gaia}, we calculate the fraction of the resolved binary systems as a function of distance.
It is plotted in red and can be explained as $k=\frac{N_b^{\rm resolved}}{N_b^{\rm resolved}+N_b^{\rm unresolved}}$.   
The corrected binary fraction for a given sample is:

\begin{gather}\label{fb2}
    f_{b}^{\rm corrected}=\frac{N_b^{\rm resolved}+N_b^{\rm unresolved}}{N_b^{\rm unresolved}+N_b^{\rm resolved}+N_s} \\
    =f_b \times \frac{1}{1-k} \nonumber
\end{gather}

where $k$ is the averaged value weighted by the distance distribution. 
Most stars in our sample have distances around 0.6 kpc (see Figure \ref{data}), so that the corrections of the resolved binary fractions is typically about 0.1 times the unresolved binary fractions.

\section{Result} \label{sec:result}

In this section, we demonstrate the results of the fitting and the bias-corrected binary fractions of the individual bin at first. The average binary fractions of the whole sample and in Galactic disks and halos are given. We also investigate the dependency of binary fraction on the chemical abundances. 

\begin{figure}[htbp] 
    \centering 
    \includegraphics[width=3in]{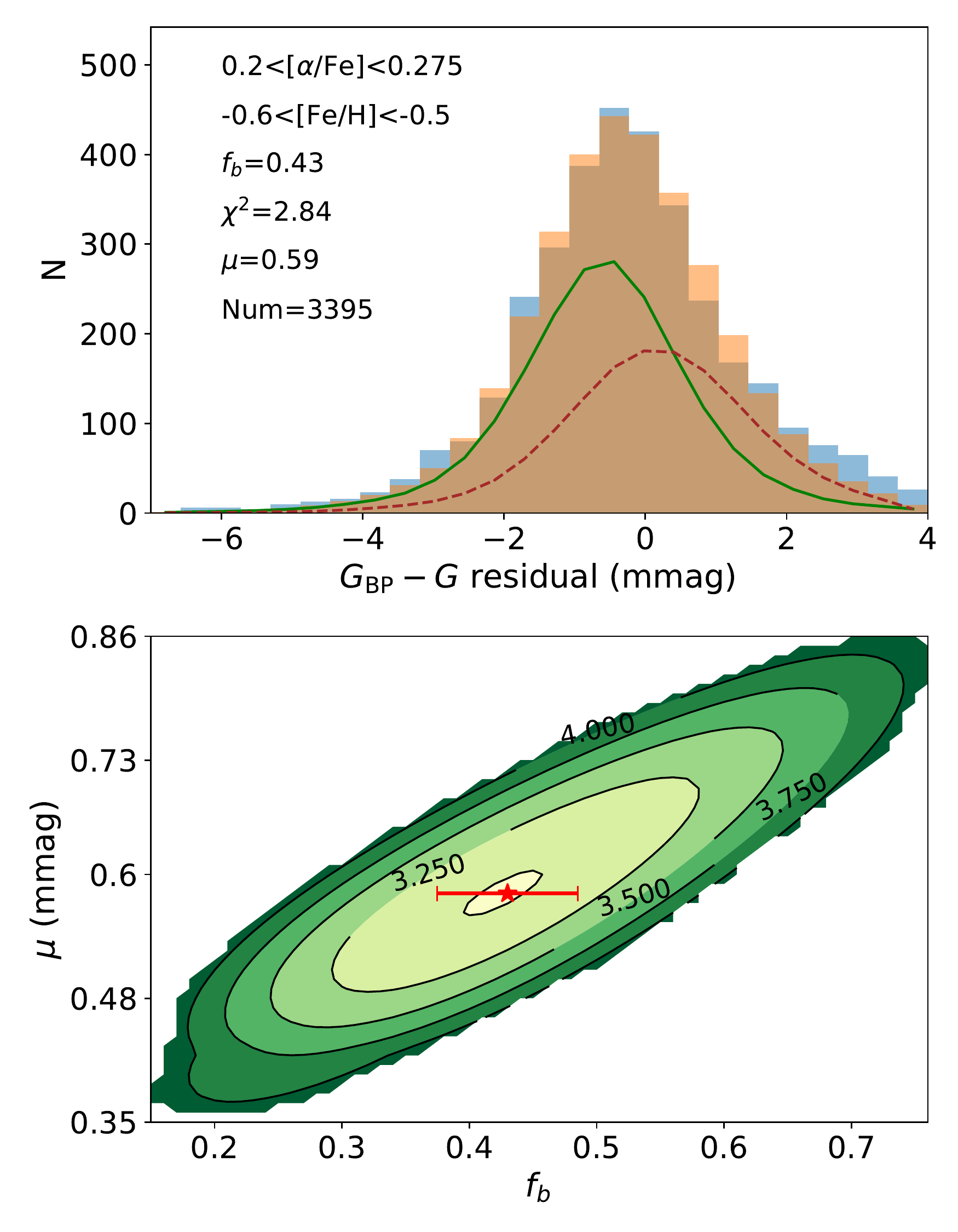}  
    \caption{Fitting result of the $-$0.6 $<$ [Fe/H] $<$ $-$0.5 and 0.2 $< [\alpha/{\rm Fe}] < $ 0.275 bin. Top panel shows the comparison between the observed and best-fitted $G_{\rm BP}-G$ residuals. Ranges of [Fe/H] and [$\alpha$/Fe], unresolved binary fraction, minimum $\chi^2$, systematic offset $\mu$, and number of stars of the bin are texted in sequence. Bottom panel shows the corresponding $\chi^2$ distribution for all possible combinations of $f_b$ and $\mu$. Red star marks the global minimum. Error bar is the formal error deduced from the Bootstrap Sample method.  \label{06histcon}} 
\end{figure}

\subsection{Binary fraction of the individual bin} \label{sec:result1}

To present the result of the best-fit model, we take one bin as an example, as shown in Figure \ref{06histcon}. The full version of the fitting result for each bin can be found in Appendix C.
In the top panel of Figure \ref{06histcon}, the blue histogram shows the distribution of the observed $G_{\rm BP}-G$ residuals. The green and red lines plot the modeled residuals of the single and binary sets, respectively. The $f_b$ and $\mu$ (in unit of mmag) are texted. Both sets have been systematically shifted toward the bluer side by the offset $\mu$. The orange histogram is the sum of the green and red lines. 
The minimum $\chi^2$, ranges of [Fe/H] and [$\alpha$/Fe], and number of stars of the bin are marked as well. We have noticed a few imperfect fitting histograms near the end of the redder side. The possible reasons and effects are discussed in the next Section. 
The corresponding $\chi^2$ distribution is shown in the bottom panel of Figure \ref{06histcon}. 
To highlight the result, we only contour the model sets with $\chi^2$ less than 1.4 times the minimum $\chi^2$. The unresolved binary fraction $f_b$ and $\mu$ are tightly correlated, the higher $f_b$, the larger $\mu$, as we expected.
As a result, the global minimum could be simply figured out through an ellipsoid fitting, resulting in a reasonably good minimum $\chi^2$ value of 2.84.
The red star donates the best-fit $f_b$ and $\mu$. The error bar of $f_b$ from the Bootstrap Sample method is over-plotted.
$f_b^{\rm corrected}$ is also subsequently obtained through Equation \ref{fb2}.

\begin{table}[htbp]
    \centering
\caption{Results of the individual bin\label{rind}}
\renewcommand\tabcolsep{2.0pt}
    \begin{tabular}{c|c|c|c|c|c|c|c}
  [Fe/H]$_l$  &  [Fe/H]$_u$  &  $[\alpha/Fe]$$_u$  &  $[\alpha/Fe]$$_l$  &  $\mu^1$ &  $f_b$ & error$^2$ &  $f_b^{\rm corrected}$\\ 
\hline
 $-$1.5 & $-$1.0 & 0.375 & 0.200 & 0.60 & 0.81 & 0.09 & 0.91 \\
 $-$1.0 & $-$0.9 & 0.375 & 0.200 & 0.57 & 0.48 & 0.13 & 0.54 \\
 $-$0.9 & $-$0.8 & 0.375 & 0.200 & 0.58 & 0.55 & 0.09 & 0.62 \\
 $-$0.8 & $-$0.7 & 0.275 & 0.200 & 0.58 & 0.43 & 0.09 & 0.48 \\
 $-$0.8 & $-$0.7 & 0.375 & 0.275 & 0.59 & 0.40 & 0.08 & 0.45 \\
 $-$0.7 & $-$0.6 & 0.200 & 0.100 & 0.71 & 0.46 & 0.13 & 0.52 \\
 $-$0.7 & $-$0.6 & 0.275 & 0.200 & 0.60 & 0.49 & 0.07 & 0.55 \\
 $-$0.7 & $-$0.6 & 0.375 & 0.275 & 0.65 & 0.51 & 0.07 & 0.58 \\
 $-$0.6 & $-$0.5 & 0.200 & 0.050 & 0.61 & 0.52 & 0.06 & 0.59 \\
 $-$0.6 & $-$0.5 & 0.275 & 0.200 & 0.59 & 0.43 & 0.05 & 0.49 \\
 $-$0.6 & $-$0.5 & 0.350 & 0.275 & 0.57 & 0.45 & 0.08 & 0.51 \\
 $-$0.5 & $-$0.4 & 0.150 & 0.050 & 0.68 & 0.39 & 0.05 & 0.44 \\
 $-$0.5 & $-$0.4 & 0.200 & 0.150 & 0.62 & 0.46 & 0.06 & 0.52 \\
 $-$0.5 & $-$0.4 & 0.275 & 0.200 & 0.58 & 0.42 & 0.05 & 0.47 \\
 $-$0.5 & $-$0.4 & 0.350 & 0.275 & 0.45 & 0.33 & 0.10 & 0.37 \\
 $-$0.4 & $-$0.3 & 0.125 & 0.050 & 0.72 & 0.35 & 0.05 & 0.40 \\
 $-$0.4 & $-$0.3 & 0.175 & 0.125 & 0.65 & 0.40 & 0.05 & 0.45 \\
 $-$0.4 & $-$0.3 & 0.225 & 0.175 & 0.58 & 0.39 & 0.06 & 0.44 \\
 $-$0.4 & $-$0.3 & 0.325 & 0.225 & 0.53 & 0.46 & 0.06 & 0.52 \\
 $-$0.3 & $-$0.2 & 0.100 & 0.025 & 0.73 & 0.30 & 0.05 & 0.34 \\
 $-$0.3 & $-$0.2 & 0.150 & 0.100 & 0.67 & 0.37 & 0.05 & 0.42 \\
 $-$0.3 & $-$0.2 & 0.200 & 0.150 & 0.59 & 0.34 & 0.07 & 0.38 \\
 $-$0.3 & $-$0.2 & 0.300 & 0.200 & 0.38 & 0.31 & 0.07 & 0.35 \\
 $-$0.2 & $-$0.1 & 0.075 & 0.000 & 0.71 & 0.27 & 0.05 & 0.31 \\
 $-$0.2 & $-$0.1 & 0.125 & 0.075 & 0.69 & 0.30 & 0.05 & 0.34 \\
 $-$0.2 & $-$0.1 & 0.175 & 0.125 & 0.74 & 0.44 & 0.08 & 0.50 \\
 $-$0.2 & $-$0.1 & 0.250 & 0.175 & 0.51 & 0.40 & 0.12 & 0.45 \\
 $-$0.1 & 0.0 & 0.050 & $-$0.025 & 0.65 & 0.25 & 0.06 & 0.29 \\
 $-$0.1 & 0.0 & 0.100 & 0.050 & 0.81 & 0.35 & 0.05 & 0.40 \\
 $-$0.1 & 0.0 & 0.175 & 0.100 & 0.61 & 0.32 & 0.10 & 0.36 \\
 0.0 & 0.1 & 0.050 & $-$0.025 & 0.73 & 0.20 & 0.07 & 0.23 \\
 0.0 & 0.1 & 0.125 & 0.050 & 0.84 & 0.25 & 0.08 & 0.28 \\
 0.1 & 0.2 & 0.050 & $-$0.025 & 0.93 & 0.26 & 0.09 & 0.30 \\
 0.1 & 0.2 & 0.125 & 0.050 & 1.10 & 0.32 & 0.13 & 0.36 \\
 0.2 & 0.3 & 0.125 & $-$0.025 & 1.19 & 0.19 & 0.16 & 0.22 \\
     \end{tabular}
 \begin{tablenotes}
     \item [1] $^1$ $\mu$ is in the unit of mmag.  
     \item [2] $^2$ Formal error deduced from the Bootstrap Sample method. 
    \end{tablenotes} 
\end{table}

\begin{figure*}[htbp] 
    \centering 
    \includegraphics[width=7in]{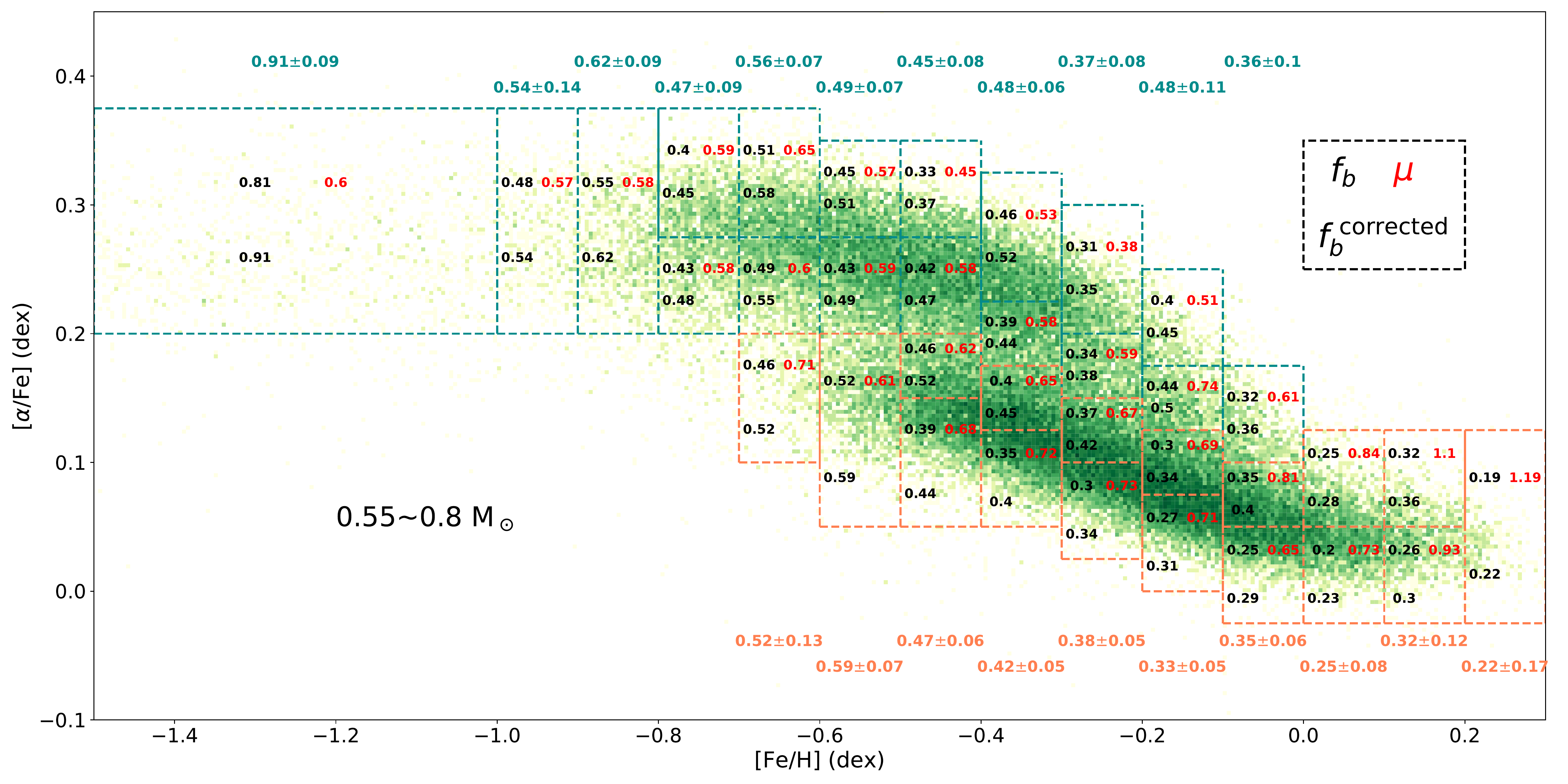}
    \caption{Summary of results for all bins. For each box, simulated unresolved binary fraction $f_b$ and bias-corrected binary fraction $f_b^{\rm corrected}$ are texted in black at the left. Offset $\mu$ in the unit of mmag is in red at the right. Orange and blue frames separately donate the thin and thick disks. The $f_b^{\rm corrected}$ and its formal error in each [Fe/H] interval is also texted in the same color.\label{062dhist}} 
\end{figure*}

We summarize the $f_b$ and the corresponding formal error, $f_b^{\rm corrected}$, and $\mu$ in Figure \ref{062dhist} and Table \ref{rind}. For each individual bin, its ranges of [Fe/H] and [$\alpha$/Fe] are illustrated by a box in Figure \ref{062dhist} and listed in Table \ref{rind}. 
In Figure \ref{062dhist}, the $f_b$, $f_b^{\rm corrected}$, and $\mu$ of each bin are texted within the box as exampled by the inset plot. 
The typical resolved binary fractions are 0.03$\sim$0.07 depending on the distance distributions. 
The 2D histograms imply the number distribution of [Fe/H] and [$\alpha$/Fe], where the thin and thick disks could be easily identified.
We use the colors of orange and blue to mark the thin and thick disks, respectively. The bin with [Fe/H] $ < -1$ is counted as the inner halo. Naturally, we calculate the average binary fractions of the thin disk, thick disk, and halo as listed in Table \ref{tab2}. 
$f_b^{\rm corrected}$ of the thick disk is clearly higher than that of the thin disk, $0.49 \pm 0.02$ versus $0.39 \pm 0.02$. It is due to the large difference of distributions of the chemical abundances in the thin and thick disks. While the binary fraction of the halo has a great probability to be larger than others. 
Notice that there are few objects in the halo in our sample, therefore the binary fraction of the halo has larger errors. 
The average binary fraction of the late G and early K type stars in the solar vicinity is $0.42 \pm 0.01$.

\begin{table}[htbp]
    \centering
    \caption{Binary Fractions of Field Stars in the Galactic Disks and Halo \label{tab2}}
    \begin{tabular}{c|c}
      & \textbf{0.55 $< m_1 <$ 0.8M$_{\odot}$} \\
    \hline
Thin Disk  & $0.39 \pm 0.02$ \\
Thick Disk & $0.49 \pm 0.02$ \\
Halo  & $ 0.91 \pm 0.09$   \\
Total  & $ 0.42 \pm 0.01$ \\
    \end{tabular}
\end{table}

\subsection{Binary fractions and chemical abundances} \label{sec:result2}

\begin{figure}[htbp] 
    \centering 
    \includegraphics[width=3.5in]{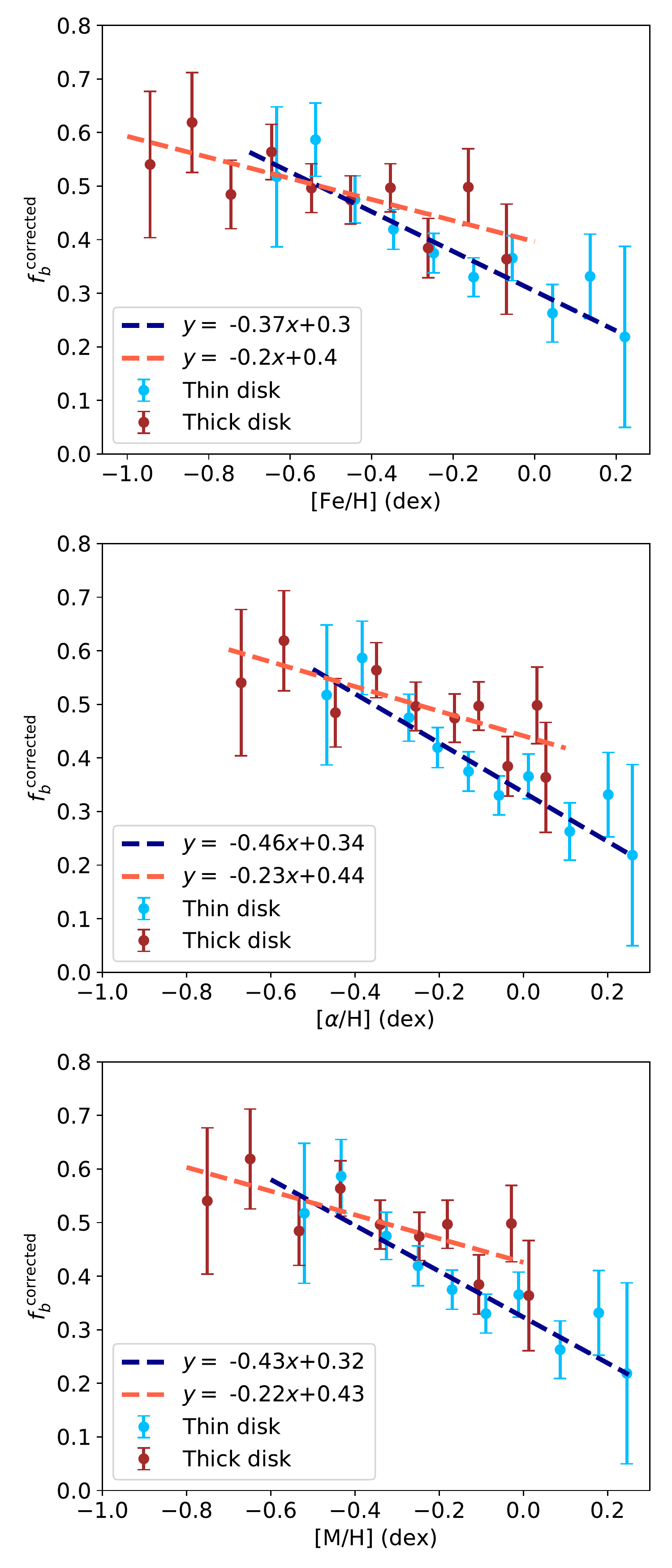}  
    \caption{Bias-corrected binary fractions as a function of [Fe/H], [$\alpha$/H], and [M/H]. In each panel, thin disk stars are in blue and thick disk stars in red. Formal errors are indicated by the vertical error bars. Linear fittings are applied and drawn in dashed lines with the slopes labelled aside. \label{bffeh}} 
\end{figure}

Current studies based on both observations (e.g., \citealp{2013araa,gao2014,2015loci2,moe2019anti,aopgee2020alpha}) and theoretical simulations (e.g., \citealp{bate2019,hurley2002}) have shown that both mass of the primary and the metallicity affect the binary fractions. 
As the mass range of the sample in this work is narrow, we suppose its impacts on the binary fractions are tiny. 
Different distributions of the chemical compositions in the Galactic disks and halo therefore lead to different average binary fractions as we have shown above.

We further check the binary fractions as a function of [Fe/H], [$\alpha$/H], and [M/H] for thin and thick disk
stars, respectively.  
The [$\alpha$/H] values are calculated by [Fe/H] and [$\alpha$/Fe]. The [M/H] values are determined according to the formula in \citet{1993Salaris} and \citet{Ferraro1999}. We separately regroup the thin and thick disk stars into ten bins in [Fe/H], [$\alpha$/H], and [M/H], with approximately equal width. The bias-corrected binary fractions as well as formal errors are deduced via the same procedures. They are scattered in Figure \ref{bffeh}, thin disk stars in blue and thick disk stars in red. 
Linear fittings to these data show that $f_b^{\rm corrected}$ roughly decreases with the increasing metallicities, consistent with the previous works (e.g., \citealp{2015loci2,moe2019anti,aopgee2020alpha}). This phenomenon is consistent with numerical simulations (e.g., \citealp{Machida2008,Machida2009,TO2014}) that 
cloud fragmentation is suppressed at higher metallicity.
The slopes of the linear fittings of thin and thick disk stars are clearly different, thin disk stars suffer a steeper slope than that of the thick disk stars, by about a factor of two.
Take the top panel as an example, thin and thick disk stars own very similar binary fractions of about 0.5 at [Fe/H] = $-$0.5 dex, the binary fraction of the thin disk stars decreases by a factor of $\sim$1.6 from [Fe/H] = $-$0.5 dex to the solar metallicity, while a factor of $\sim$1.2 for the thick disk stars. 
Tendencies along with the [$\alpha$/H] and [M/H] show similar results in Figure \ref{bffeh}. 

Thin and thick disk stars are believed to have different formation histories. 
In particular, $\alpha$-elements enrichments are mainly done by the core-collapse supernovae explosions, which
happened earlier than the iron enrichments by the type Ia supernovae explosions. 
Thick disk stars were formed on a short timescale when little iron was contained in the interstellar medium. Whereas thin disk stars were constantly formed in the late time (e.g., \citealt{2002ARAA} and references therein).
Therefore the formation and evolution process of the thin and disk stars could be different, consequently the binary fractions.

\begin{figure*}[htbp] 
    \centering 
    \includegraphics[width=6.5in]{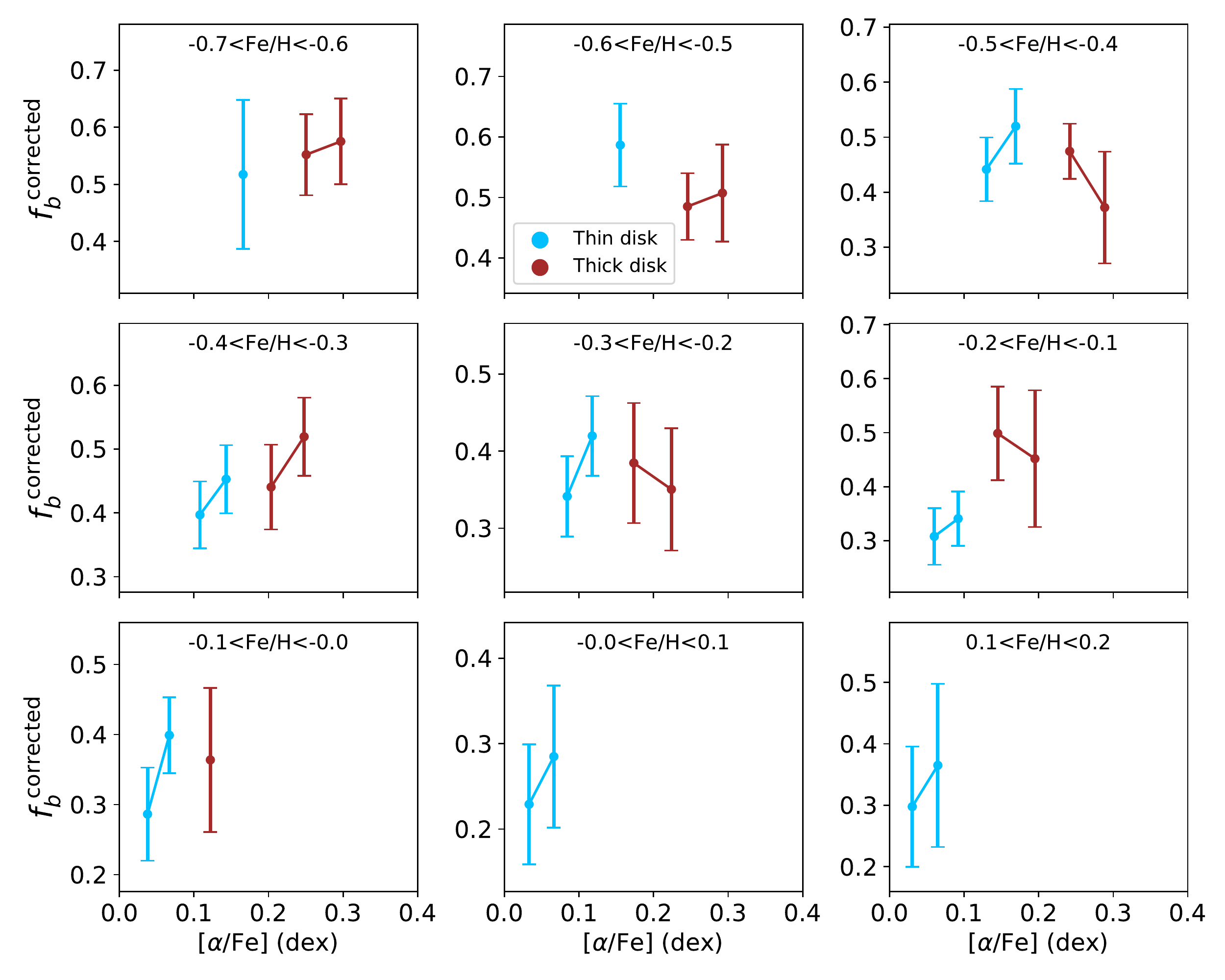}  
    \caption{Binary fractions against [$\alpha$/Fe] at every [Fe/H] interval of 1 dex. Bins from the thin and thick disks are respectively scattered in blue and red dots. \label{bfalfe}} 
\end{figure*}

Another difference between the thin and thick disk stars is how the binary fractions vary with [$\alpha$/Fe] for a given [Fe/H] value. 
For every 0.1 dex of [Fe/H] in Figure \ref{062dhist}, we select samples having two [$\alpha$/Fe] bins within either thin disk or thick disk and plot the $f_b^{\rm corrected}$ against [$\alpha$/Fe] in Figure \ref{bfalfe}. The [$\alpha$/Fe] values are the median values of all stars included in each bin. Same as Figure \ref{bffeh}, blue dots are the thin disk stars, red dots are the thick disk stars. We generate a Gaussian likelihood distribution of the binary fraction for each point taking the formal error into account. By comparing the likelihood distributions of two bins of one [Fe/H] interval within either thin disk or thick disk, we obtain the possibility that stars having larger [$\alpha$/Fe] own the larger binary fraction.
For the thin disk, from metal poor to metal rich, the possibilities are 0.81, 0.77, 0.85, 0.68, 0.91, 0.70, and 0.66, respectively. But for the thick disk, this tendency is not clear.
A rough estimation of the typical [$\alpha$/Fe] impacts is deduced for thin disk stars. 
For a given [Fe/H], 0.05 dex of increasing [$\alpha$/Fe] raises the binary fractions by about 0.1. 
We could connect the above phenomenon with the stellar age distributions in the [Fe/H] and [$\alpha$/Fe] panel.
Plenty works (e.g., \citealp{2001Matteucci,agemetal,2019Feuillet,huang2020}) have demonstrated that stars of the high [$\alpha$/Fe] sequence typically have ages older than 9--10 Gyr, showing a nearly flat distribution of ages for a given [Fe/H] value. In contrast, a certain gradient exists in the generally young ($<$8 Gyr) populations, with older stars showing higher [$\alpha$/Fe] values. We suppose that both the binary fractions and ages serve the hints of the different formation histories.

\section{Discussion} \label{sec:dis}

\subsection{Redder excess of the color residual}

In this paper, when fitting the binary effect on the $G_{\rm BP}-G$ residuals, the lower and upper limits of the $G_{\rm BP}-G$ residuals are respectively $-7$ and 4 mmag. 
Fractions of the objects beyond these boundaries are typically 0.3 per cent for $G_{\rm BP}-G$ residuals less than -7 mmag and 5 per cent for $G_{\rm BP}-G$ residuals greater than 4 mmag. 
Reasons of the excess of the observing residuals in Figure \ref{06hist} include:
(1) underestimated photometric errors for few stars in \textit{Gaia};
(2) contamination from close foreground or background objects; and 
(3) flux excess due to the unresolved multiple systems. 
The first effect should be symmetric, it barely affects the fitting results since the ignoring fractions are tiny. 
The second and third effects are expected to be similar with the binary effect. LAMOST has applied elaborate target selections to avoid the second situation. We also restrict the sample to the high galactic latitude stars, there are little chances of the second situation. The third case is worthy to be discussed.

It is hard to give reliable corrections on the binary fractions caused by the effect from multiple systems. 
Firstly, it is unpractical to model the various structures of multiple systems. The mass ratio distributions of the short- and long-period subsystems are different \citep{raghavan2010,2008Tokovinin,2012bate}. The period distributions of the subsystems also vary with different kinds of hierarchies \citep{2014Tokovinin}. 
Secondly, the missing multiple fraction is uncertain. Although the fraction of multiple systems is reported to be 0.11--0.17 for the solar-mass stars in general (\citealp{raghavan2010,dm91,2012bate,2014Tokovinin1}), it changes with the orbital period \citep{2012allen,2008Makarov} in a complex way. 
Notice that a substantial portion of the multiple systems contain a very low-mass third/fourth component, whose contribution to the color residuals can be ignored. Therefore, such systems have been well accounted in our results.

\subsection{Comparison with previous works}\label{com}

Comparing to previous works based on other approaches, our method is weakly limited by the binary orbital period. Summarized illustrations of the ranges of binary periods and mass ratios that can be detected with different methods are shown in \citet{moe2017} and \citet{2018specbi}.
For instance, the radial velocity monitoring method and light curve analysis method (for eclipsing binaries) may only work for close binaries.
The former is suitable for log$_{10}$($P/{\rm day}$) $\leq 3$, while the later works only at even a shorter period of log$_{10}$($P/{\rm day}$) $\leq 2$. The common proper motion and visual binaries are only sensitive to wide binaries.
The spectral fitting method could achieve log$_{10}$($P/{\rm day}$) $\leq 8$ but only for the binaries with intermediate mass ratio.

In addition to periods, these methods also suffer from other specific limitations. 
Taking several examples, radial velocity monitoring technique requires time-domain data. Photometric technique is confined to the systems with high inclination angles. Astrometric techniques like from \textit{Hipparcos} and \textit{Gaia} are biased to the bright binaries having intermediate separations based on the angular resolution and time baseline of the surveys. 
As for the SLOT method, its realization is solely taking advantage of the difference between the color deduced by the metallicity-dependent stellar loci (of the single star) and the combined color (of binary system). It covers both close and wide binaries, supporting log$_{10}$($P/{\rm day}$) less than 6--8 depending on the distance distribution and instrument spatial resolution. Besides, it depends weakly on the mass ratio distribution and can be carried out with single epoch data.

We stress that most previous studies on binary fraction are restricted to the thin disk stars due to the selection bias. 
It would be reasonable to compare their results with ours of the thin disk. 
For example, the G2–K3 sample in \citet{raghavan2010} has a binary fraction of $0.41 \pm 0.03$, 
consistent with with our $0.39 \pm 0.02$. 
Our work basically uses the the same method as \citet{2015loci2} except for taking the resolved binaries into consideration. 
Our sample is also comparable with the $0.9 < g-i < 1.2$ mag and $-1 <$ [Fe/H] $< 0$ sample in \citet{2015loci2}. Their binary fraction is 0.35, in good agreement with the average fraction of the unresolved binary 0.37 in this work.

Comparing with \citet{aopgee2020alpha}, where the close binary fraction decreases by a factor of $\sim$2.4 from [Fe/H] = $-$0.5 dex to [Fe/H] = 0.25 dex, our binary fraction decreases by a factor of $\sim$2.3 from [Fe/H] = $-$0.5 dex to [Fe/H] = 0.25 dex. 
The decreasing factors seem to be in good agreement.
We adopt a universal period distribution for binaries with different metallicities. 
However, recent works (e.g., \citealp{TO2014}, \citealp{Jayasinghe2020}) indicate that the binary period shifts toward smaller separations as the metallicity decreases. 
If it is true, there would be an over-correction of binary fraction for the resolved metal-poor binary. 
We make a simple estimation that a log-normal Gaussian distribution with a mean of log$P$ = 4 and $\sigma_{{\rm log}P}$ = 1.5 for [Fe/H] $<$ $-$0.2 reduces the binary fraction by $\sim$0.045\footnote{\citet{aopgee2020alpha} made the same simulation and found no significant change from their published results.}. Consequently, it would lead to a slower slope of the metallicity dependency. 
Given that our sample also contains wide binaries in addition to the close binaries, this could be accounted for that the metallicity dependency of the binary fraction changes with the increasing period (e.g., \citealp{widedr2,widefb2021,bate2019}). As the period increases, the anti-correlation between the close binary fraction and metallicity weakens, even becomes positive.

One motivation of this paper is exploring the binary fraction as a function of the $\alpha$ enrichment. 
Figure \ref{bfalfe} suggests that they are most likely positively correlated. 
This result is not affected by the assumed period distribution, as it is under the condition of a given [Fe/H].
\citet{aopgee2020alpha} take the opposite view. The indicator used in their paper is $[\alpha$/H], which can be easily converted to [$\alpha$/Fe] for a given [Fe/H] value. 
Their result that the anti-correlation between $\alpha$ abundances and multiplicity is mainly concluded from the $\alpha$-poor sample, like $-0.2 < [\alpha$/H] $< 0.1$ at solar [Fe/H] (see their Figure 10). 
The result could be affected by the wider mass ranges (0.5 -- 2.0 M$_{\odot}$) of their sample compared to our sample (0.55 -- 0.8 M$_{\odot}$). On one hand, high-mass stars are generally younger than low-mass stars, and subsequently have lower $\alpha$ abundance. On the other hand, 
the close binary fraction increases by a factor of 2 from 0.5 to 2 solar mass \citep{moe2017}.
Therefore, the higher multiplicity of the lower $\alpha$ abundance stars could be a joint result of the higher mass and lower $\alpha$ abundance instead of lower $\alpha$ abundance alone. 
They do find that the close binary fraction flattens for [$\alpha$/Fe] $>$ 0.05 dex.
Moreover, in their Figure B3, for 0.125 $<$ [Fe/H] $<$ 0.275, the binary fraction increases with the increasing [$\alpha$/H] at [$\alpha$/H] $>$ 0.1, consistent with the last panel of our Figure \ref{bfalfe}.

\section{Summary} \label{sec:sum}

In this work, the binary fraction of individual stellar population is derived using the SLOT method and corrected for the effect of resolved binaries. 
The SLOT method is described in detail in Section \ref{sec:method}. Compared with other methods, such as radial velocity monitoring and light curve analysis, our method suffers from less limitation on the period, mass ratio, and inclination, etc.

We use a distance-limited sample of 90 thousand late G and early K dwarf stars selected from \textit{Gaia} EDR3 and LAMOST DR5 data, in which \textit{Gaia} photometry has been delicately re-calibrated by \citet{niudr3} and \citet{yl}. 
Our sample covers wide ranges in chemical abundances, especially for $\alpha$ abundances comparing with the previous studies. 
We divide the sample into different bins based on their [Fe/H] and [$\alpha$/Fe] values. Results of the individual bins are demonstrated in Subsection \ref{sec:result1}. All bins are classified into the thin disk, thick disk, and halo when exploring the tendencies of the binary fractions.
The average binary fraction of the whole sample is $0.42 \pm 0.01$. The thin disk stars 
are found to have a lower binary fraction ($0.39 \pm 0.02$) than the one of the thick disk stars ($0.49 \pm 0.02$).
The halo stars have the highest binary fraction ($0.91 \pm 0.09$). Our estimations of the binary fractions are consistent with previous studies \citep{raghavan2010,2015loci2}.

As shown in Subsection \ref{sec:result2}, we find that the binary fraction is not just simply anti-correlated with [Fe/H], [$\alpha$/H], and [M/H]. The impacts of chemical abundances are different for thin and thick disk stars.
It is shown that (1) metallicities depress the binary fractions more significantly in the thin disk than in the thick disk; (2) for a given iron abundance, the binary fractions increase with the increasing $\alpha$ abundance in the thin disk, 
the trend disappears in the thick disk.
In particular, the slope of the binary fraction as a function of [Fe/H] alone is $-$0.37 per dex for the thin disk stars and $-$0.2 per dex for the thick disk stars. The impacts of [$\alpha$/H] and [M/H] exhibit similar trends. 
Also, given a [Fe/H], [$\alpha$/Fe] increase by 0.05 dex, the thin disk binary fractions typically increase by 0.1. 
These might be the consequences of the different formation histories of the thin and thick disks.
Our results provide new clues for theoretical works on binary formation.

A single-power-law is assumed for the binary mass ratio distribution in this work. 
In the future, by using distance information, we are going to determine the mass ratio distribution and the binary fraction at the same time.

\acknowledgments
We acknowledge the anonymous referee for his/her valuable comments that improve the quality of this paper significantly.
We acknowledge helpful discussions with Prof. Liu Chao, Zhang Haopeng, and Wang Yilun. This work is supported by National Science Foundation of China (NSFC) under grant numbers 12173007, 11603002, 11988101, and 113300034, National Key Research and Development Program of China (NKRDPC) under grant numbers 2016YFA0400804, 2019YFA0405503, and 2019YFA0405504. 
S. W. acknowledges support from the Youth Innovation Promotion Association of the CAS (id. 2019057).
This work has made use of data products from the Guoshoujing Telescope (the Large Sky Area Multi-Object Fiber Spectroscopic Telescope, LAMOST). LAMOST is a National Major Scientific Project built by the Chinese Academy of Sciences. Funding for the project has been provided by the National Development and Reform Commission. LAMOST is operated and managed by the National Astronomical Observatories, Chinese Academy of Sciences. This work has made use of data from the European Space Agency (ESA) mission {\it Gaia} (\url{https://www.cosmos.esa.int/gaia}), processed by the {\it Gaia} Data Processing and Analysis Consortium (DPAC, \url{https://www.cosmos.esa.int/web/gaia/dpac/consortium}). Funding for the DPAC has been provided by national institutions, in particular the institutions participating in the {\it Gaia} Multilateral Agreement.

\appendix
\setcounter{table}{0}   
\setcounter{figure}{0}

\renewcommand\thetable{\Alph{section}\arabic{table}} 
\section{Remove MSTO}
The MSTO stars are excluded from our sample as shown in Figure \ref{turnoff}. They are selected following the criterion $M_{\rm G} < a \times (G_{\rm BP}-G_{\rm RP}) +b$, where $a$ is $-$3.57 for all [Fe/H] bins and $b$ are listed in Table \ref{taba1}. 

\begin{table}[htbp]
\centering
\caption{Criteria of removing MSTO stars}\label{taba1}
\begin{tabular}{c|cccccccc}
[Fe/H] & $-$1.5  & $-$1.25 & $-$1.   & $-$0.75 & $-$0.5  & $-$0.25 &  0.   &  0.25 \\
\hline
$b$ & 6.7 & 6.9 & 7.1 & 7.3 & 7.5 & 7.6 & 7.7 & 7.8 \\
\end{tabular}
\end{table}

\renewcommand\thefigure{\Alph{section}\arabic{figure}} 
\section{PARSEC model}

\begin{figure*}[htbp] 
    \centering 
    \includegraphics[width=6.5in]{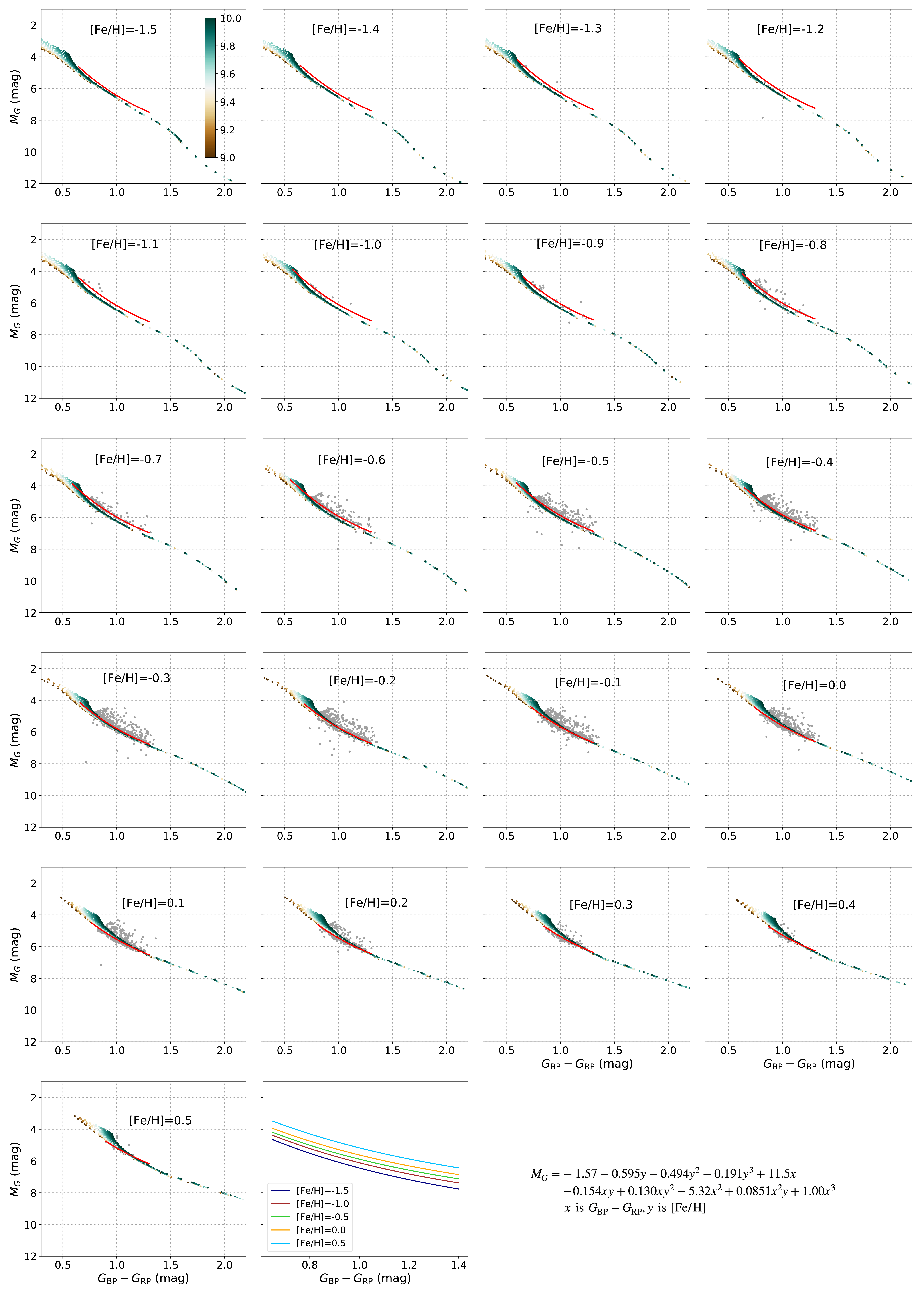}  
    \caption{\textit{grey}: dwarf samples used in this work. \textit{red lines}: fitted $M_G=f(G_{\rm BP}-G_{\rm RP},{\rm [Fe/H]})$ relation as texted in the figure. \textit{colored tracks}: Color-magnitude model of the PARSEC isochrones used in this work. Ages are indicated in the color bar. Metallicities are labelled in the top center of each panel. Variations of the relation of different metallicities are displayed in the last panel. \label{cmd}} 
\end{figure*}

The PARSEC model is applied in this work when calculating the stellar mass and the absolute $G$ magnitude. We extract isochrones of different sets of [Fe/H] with step of 0.1 dex from $-$1.5 to 0.5 dex and age with step of 1 Gyr from 1 to 10 Gyr. In Figure \ref{cmd}, grey points are the sample we used, colored tracks are the color-magnitude tracks from the PARSEC model. Red lines that laying on the PARSEC model are empirically fitted using the grey points, in which 2-sigma clipping is performed. Some grey points scatter above the red line as well as the tracks are binaries. Our sample is in good agreement with the PARSEC model. We can see that the width of tracks caused by ages among the color ranges of our sample is narrow. Hence we just interpolate the tracks to have the $M_{\rm G}$. 

\begin{figure*}[htbp] 
    \centering 
    \includegraphics[width=6.5in]{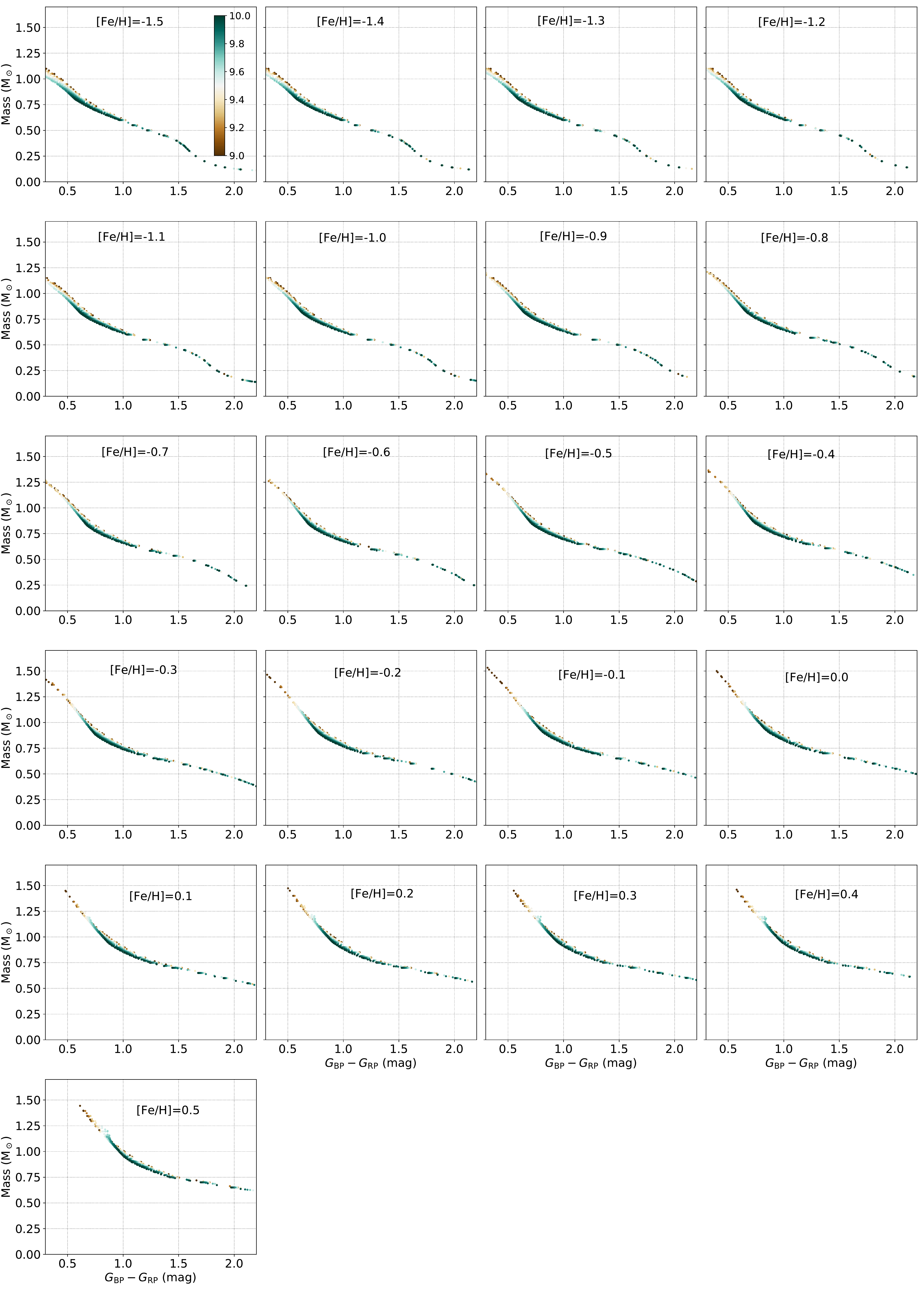}  
    \caption{Color-mass model of the PARSEC isochrones used in this work. Ages are indicated in the color bar. Metallicities are labelled in the top center of each panel. \label{cm}}
\end{figure*}

Figure \ref{cm} is the color-mass model of the PARSEC model. Stellar mass is derived in the same way as the $M_{\rm G}$.

\renewcommand\thefigure{\Alph{section}\arabic{figure}} 
\section{Fitting result of the individual bin}

\begin{figure*}[htbp] 
    \centering 
    \includegraphics[width=6.5in]{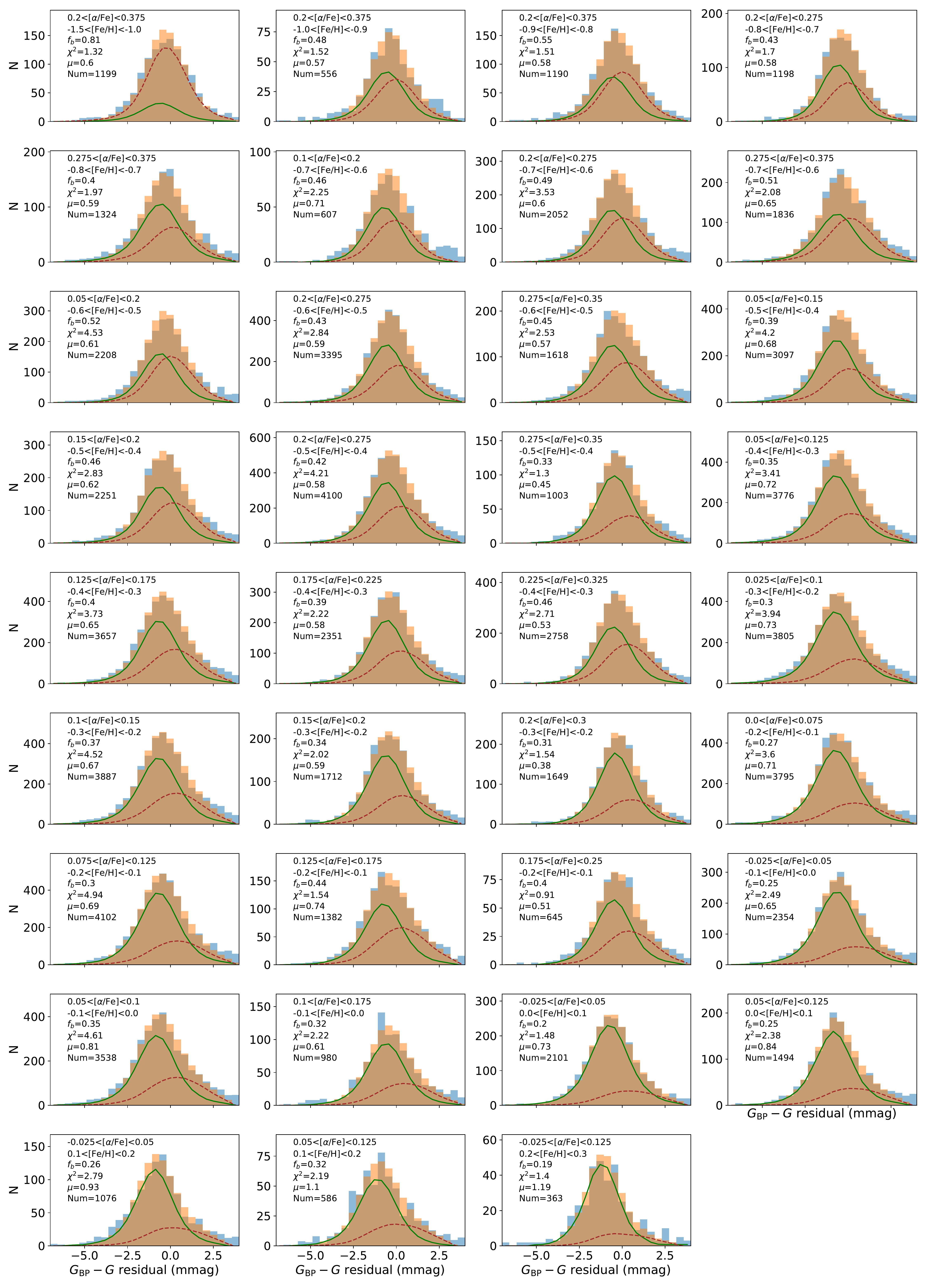}  
    \caption{Same to the top panel of Figure \ref{06histcon} but for all bins of [Fe/H] and [$\alpha$/Fe].\label{06hist}} 
\end{figure*}

\begin{figure*}[htbp] 
    \centering 
    \includegraphics[width=6.5in]{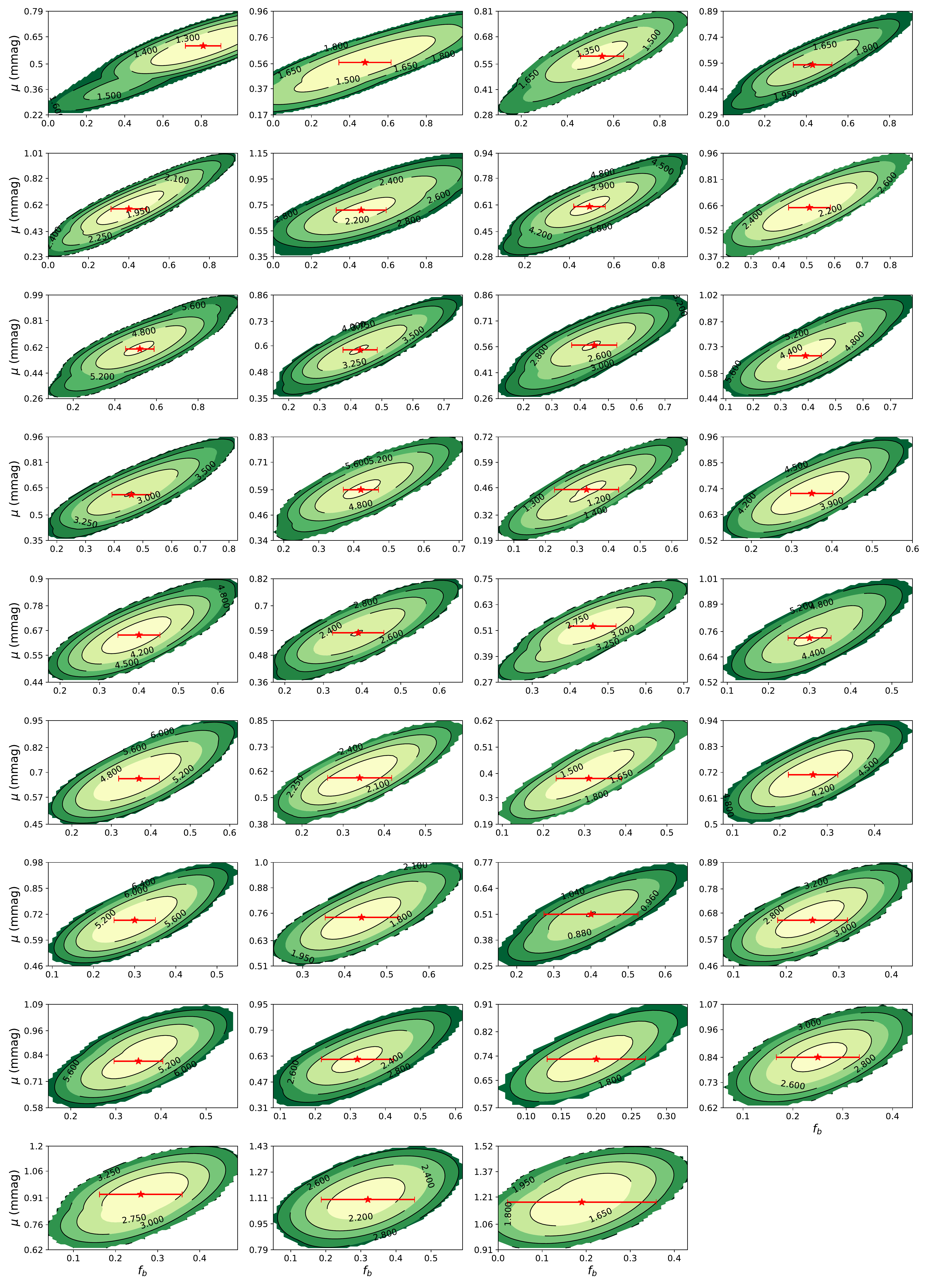}  
    \caption{Same to the bottom panel of Figure \ref{06histcon} but for all bins of [Fe/H] and [$\alpha$/Fe].\label{06con}} 
\end{figure*}

Same as Figure \ref{06histcon}, the fitting results of all bins are shown in Figure \ref{06hist} and \ref{06con}.
The typical minimum $\chi^2$ values are between 1.0 -- 5.0. 
Almost all $\mu$ values are less than 1 mmag. Despite the existence of binary stars in the sample when fitting the loci, our stellar loci listed in Table \ref{tab1} are reasonably accurate for almost all cases.

\bibliography{sample63}{}
\bibliographystyle{aasjournal}

\end{document}